\documentclass[]{aastex631}

\usepackage{booktabs}

\newcommand{\lf}[1]{{\color{black}#1}}
\newcommand{\smm}[1]{{\color{black}#1}}
\newcommand{\ya}[1]{{\color{black}#1}}
\usepackage{soul}
\usepackage{cancel}
\usepackage{multirow}
\usepackage{parskip}
\usepackage{appendix}

\graphicspath{{./}{figures/}}
\begin{document}

\title{Enhanced three-minute oscillation above a sunspot during a solar flare}

\correspondingauthor{Ya Wang}
\email{wangya@pmo.ac.cn}

\author[0000-0003-3699-4986]{Ya Wang}
\affiliation{Key Laboratory of Dark Matter and Space Astronomy, Purple Mountain Observatory, CAS, Nanjing, 210023, People's Republic of China \\
}
\affiliation{SUPA School of Physics \& Astronomy, University of Glasgow, G12 8QQ,UK \\
}

\author[0000-0001-9315-7899]{Lyndsay Fletcher}

\affiliation{SUPA School of Physics \& Astronomy, University of Glasgow, G12 8QQ,UK \\
}

\affiliation{Rosseland Centre for Solar Physics, University of Oslo, P.O.Box 1029 Blindern, No-0315 Oslo, Norway \\
}

\author[0000-0002-9242-2643]{Sargam Mulay}
\affiliation{SUPA School of Physics \& Astronomy, University of Glasgow, G12 8QQ,UK \\
}

%\collaboration{20}{(AAS Journals Data Editors)}

\author[0000-0002-5898-2284]{Haisheng Ji}
\affiliation{Key Laboratory of Dark Matter and Space Astronomy, Purple Mountain Observatory, CAS, Nanjing, 210023, People's Republic of China \\
}

\affiliation{School of Astronomy and Space Science, University of Science and Technology of China, Hefei, Anhui, 230026, People's Republic of China\\
}

\author[0000-0003-2427-6047]{Wenda Cao}
\affiliation{Big Bear Solar Observatory, New Jersey Institute of Technology, Big Bear City, CA 92314, USA \\
}




%%%%%%%%%%%%%%%%%%%%%%%%%%%
\begin{abstract}

%Solar flares are one of the most powerful magnetic events on the Sun. In this paper, we continue to report the M-class flare on 2012 July 5 in the active region (AR) 11515. We concentrate on the oscillation of the footpoint of the late phase loop as observed by the high-resolution imaging at H$\alpha$ 6563~\AA\ and He~\textsc{i} 10830~\AA\ from the Goode Solar Telescope (GST) at the Big Bear Solar Observatory (BBSO) and the Atmospheric Imaging Assembly (AIA) on board the Solar Dynamic Observatory (SDO). One of the footpoints of the late phase loop roots in the boundary of the sunspot's umbra and penumbra. It oscillates for a period of three minutes, the oscillation is observed from the lower atmosphere to the corona. This provides a piece of clear evidence that the three-minute oscillation of the footpoint of the late phase loop above the chromospheric umbra, presenting that the magnetoacoustic wave is strengthened by the pulse from the magnetic reconnection which leads to extra heating on the late phase loop or the magnetoacoustic wave modulates the magnetic reconnection occurred during the late phase. As well as, the magnetoacoustic wave propagates upwardly from the chromosphere into the corona. On the other hand, before and after the flare main phase, the running penumbral wave turns into oscillation at the footpoint. The time interval of the running penumbral wave is about 4 minutes, the slight difference in the frequency before and after the main phase mainly depends on the inclination angle $\theta$ of the magnetic field.

\lf Three-minute oscillations are a common phenomenon in the solar chromosphere above \lf{a} sunspot. Oscillations can be affected by the energy release process related to solar flares. 
In this paper, \lf{we report on an enhanced oscillation in flare event SOL2012-07-05T21:42 with a period of around three minutes, that occurred at the location of a flare ribbon at a sunspot umbra-penumbra boundary, and was observed both in chromospheric and coronal passbands}. An analysis of this oscillation was carried out using simultaneous ground-based \smm{observations} from \smm{the} Goode Solar Telescope (GST) at the Big Bear Solar Observatory (BBSO) and space-based \smm{observations} from \smm{the} Solar Dynamics Observatory (SDO). \lf{A} frequency shift was observed before and after the flare, \lf{with} the running penumbral wave \lf{that was present} with a period of \lf{about 200~s} before the flare \ya{co-existing with a} strengthened oscillation with a period of \lf{180~s} at the same locations after the flare. We also found a phase difference between different passbands, \lf{with} the oscillation \lf{occurring} from high-temperature to low-temperature passbands.
\lf{Theoretically,} the change in frequency is strongly dependent on the variation of the inclination of the magnetic field and the \lf{chromospheric temperature}. \lf{Following an analysis of the properties of the region,} 
\ya{we find the frequency change is caused by the slight decrease of the magnetic inclination angle to the local vertical. In addition, we suggest that the enhanced three-minute oscillation is related to the additional heating, maybe due to the downflow, during the EUV late phase of the flare.} 

%, which may be a result of the standing wave in the late-phase loop. 

\end{abstract}
%%%%%%%%%%%%%%%%%%%%%%%%%%%

%% You will be asked to select these concepts during the submission process


%%%%%%%%%%%%%%%%%%%%%%%%%%%
\section{Introduction} 
\label{sec1:introduction}

Sunspot oscillations have been observed and studied for decades \citep{1969SoPh....7..351B}. Three types of oscillations have been reported: the five-minute oscillation at the photosphere \citep{1970ApJ...162..993U, 1989MmSAI..60...71M}, running penumbral \lf{waves} \citep{1974SoPh...38..399N, 2018ApJ...852...15P}, and three-minute oscillations in the chromosphere above sunspots \citep{2006RSPTA.364..313B}. Five-minute oscillations are acoustic-type vibrations (p-modes), the origin of which is believed to be the standing acoustic waves in a subphotospheric cavity \citep{1970ApJ...162..993U}. \lf{They are used to probe} the solar interior. Observation shows that five-minute oscillation with reduced amplitude in the umbra acts as a filter in transmitting selected frequencies \citep[]{1986ApJ...311.1015A}. \citet{1988ApJ...335.1015B} reported that the power of p-mode oscillation has been absorbed in sunspots and the lifetime of high-degree p-mode may be reduced during solar activities. The spectroscopy of sunspots investigated the power distribution for the frequency in a sunspot and the velocity power spectra showed a decrease in the range of five minutes in the sunspot \citep {1996A&A...315..603B}. Observations presented by \citet{2004A&A...424..671K} also show the five-minute p-mode oscillations propagate across the entire sunspot. \lf{Five-minute oscillations in fact} show a broad distribution of frequencies, with a peak power around a period of roughly five minutes. \\%The great success of the concept of trapped subphotospheric oscillations marked the beginning of helioseismology. (references)\\

A significant phenomenon in the sunspot penumbra is the \lf{r}unning \lf{p}enumbral \lf{w}ave (RPW). The running penumbral wave describes chromospheric H$\alpha$ velocity and intensity fronts that \lf{are} observed moving out through the sunspot penumbra \citep{1972ApJ...178L..85Z}. %The running penumbral wave is a periodic, outwardly propagating wave that propagates through the penumbra of a sunspot (Hirzberger et al., 2002). At the umbra-penumbra boundary, phase speeds may be as large as 15-20~{$\rm km\,s^{-1}$}, and they may decrease to 4-7~{$\rm km\,s^{-1}$}.
 \lf{Based on their visual pattern,} many authors provided evidence that the running penumbral wave is caused by field-aligned acoustic waves propagating up in the sunspot \citep[e.g.][]{2007ApJ...671.1005B}. In this scenario, the pattern of delayed wavefronts gives rise to the apparent outward motion of RPW, which may also explain the large range of wave speeds. This scenario \lf{also} indicates that the RPW can occur at the edge of \lf{a} pore, and the existence of the penumbra is not necessary. \citet{2015ApJ...802...45C} provided the observed wave properties of pores, which are similar to \lf{those in} sunspots, by using the Ca~\textsc{ii} 8542~\AA\ and H$\alpha$ lines. The results supported the explanation of the observed wave as a slow magnetoacoustic wave propagating along the magnetic field in pores. The apparent horizontal motion can be explained by the projection effect caused by the inclination of the magnetic field with a canopy structure. \\

%These waves are believed to be caused by magnetohydrodynamic instabilities that propagate along magnetic field lines of the penumbra. Further research has identified two types of running penumbral waves: outward- and inward-propagating. According to research studies, outward-propagating running penumbral waves were first reported by \citet{scharmer2002penumbral}, and later studied in detail by \citet{rimmele2008filaments}. Inward-propagating running penumbral waves were discovered by \citet{westendorp2001magnetic} and \citet{wiehr2004wave}. Both types of penumbral waves are found in sunspots, usually associated with a photospheric horizontal magnetic field component parallel to the penumbral filaments \citep{bellot2005solar, zakharov2008longitudinal, rajaguru2016asymmetry}. The former has been observed to move at speeds of up to 35~{$\rm km\,s^{-1}$}, while the latter has been found to move at an even higher speed of 60~{$\rm km\,s^{-1}$}. Meunier et al. (2007) reported that the mean period was 8.7 ± 0.3 minutes and showed no dependence on the tilt angle of the penumbrae. Subramanian et al. (2008) suggested that the period increases with increasing azimuthal angle. Yurchyshyn et al. (2008) found that the period varies between 5 and 10 minutes, with an average of around 6 minutes.\\

In this study, we mainly focus on the \lf{third type, the} three-minute oscillations \citep{1991A&A...250..235F, 2019A&A...627A.169F}. It is generally believed that three-minute oscillations are slow magneto-acoustic waves. %A 
 For a small plasma $\beta$, \smm{where \(\beta = 8\pi p/B^{2}\ll1\)} the slow magneto-acoustic oscillation \lf{disturbs the magnetic field very little}, and its behaviour is the same as that for sound \lf{waves}. In the low-$\beta$ plasma in a sunspot, slow magneto-acoustic waves are field-aligned compressive motions of the plasma, moving with the local sound speed along the magnetic field. Since the Alfv\'en speed exceeds the sound speed throughout the optically thin umbral atmosphere, \lf{and taking into account the solar gravity,} these fluctuations are classified as low-$\beta$ slow magneto-acoustic-gravity (MAG) waves \citep{1977A&A....55..239B}. \lf{The three-minute oscillation is generally considered as being due to the leakage of the higher frequencies in the five-minute oscillation spectrum. Waves with frequencies higher than the local acoustic cut-off frequency are able to move upwards from the photosphere through the chromosphere into the corona}.  %In this scenario, the wave originates from the deeper layers of the solar atmosphere and propagates upward along the magnetic field lines. 
 For example, \cite{2006ApJ...640.1153C} reported \lf{that} the three-minute power observed at chromospheric heights comes directly from the photosphere by means of linear wave propagation in between the levels of formation of the silicon and helium lines by using full Stokes vector IR spectro-polarimetry. \\

 Another explanation is that the chromosphere is  an acoustic resonator, \lf{with a cavity formed} between the photosphere and the transition region, which partially reflects the slow magneto-acoustic wave \citep[e.g.,][]{2008SoPh..251..501Z, 2011ApJ...728...84B}. In this theory, the parts of the solar p-mode spectrum with frequencies  %\lf{\sout{three-minute oscillations are excited by}} disturbances 
 %\lf{\sout{outside or}} below the sunspot with frequencies 
 equal to or greater than the acoustic cut-off frequency transmit into the chromospheric cavity, which resonates  at the acoustic cut-off frequency. \lf{However, as it} is a leaky resonator, the oscillations can propagate upward into the corona. The acoustic cut-off frequency is determined by \smm{\(\omega \propto cos~\theta/\sqrt{T}\)} \citep{1991A&A...250..235F}, where \(\theta\) is the angle between the magnetic field and the vertical, and \(T\) is the plasma temperature. Different temperature profiles of the sunspot's umbra would lead to different peaks in the spectrum of the chromospheric resonator, explaining the frequency variation in the sunspot oscillation. \lf{The existence of a chromospheric resonator is a matter of some debate \citep{2020NatAs...4..220J, 2020A&A...640A...4F, 2021A&A...645L..12F, 2021NatAs...5....5J}} but opinion seems to be moving in its favour. Recent numerical modelling by \cite{2020A&A...640A...4F} reveals that different profiles of the \lf{chromospheric} temperature and density lead to \lf{variations in the cutoff frequency (that are quite different from analytic model predictions) so we would expect this to affect the observed frequency of oscillation. These authors also looked at the effect of changing the strength and inclination of the sunspot magnetic field}. There is evidence of variation in the three-minute oscillation spectrum across a sunspot. The frequency decreases in the horizontal direction from the sunspot center, due to the field lines being almost vertical in the center with gradually increasing inclination towards the edge of the sunspot \citep{2014A&A...569A..72S}. \lf{The $cos~\theta$ dependence of the cutoff frequency was verified numerically by \cite{2020A&A...640A...4F} in regions where the temperature gradient was not too extreme.}\\

On the other hand, flare \lf{q}uasi-\lf{p}eriodic \lf{p}ulsations \smm{(QPP)} \citep{2009SSRv..149..119N, 2019ApJ...875...33H, 2021SSRv..217...66Z} are a frequent phenomenon which has been known for over 50~years. Flare QPPs are defined as a sequence of bursts of flare emission with similar time intervals between successive peaks. Typical periods of QPPs are in the range of a few seconds to a few minutes. %The QPPs property expands to the stellar flares based on solar-stellar analogies. 
It has been found that QPPs can appear in \smm{various} phases of a solar flare, \smm{such as} the preflare phase \citep{2020A&A...639L...5L}, impulsive phase, and gradual phase \citep{2021SSRv..217...66Z}. \smm{Various} QPPs mechanisms in solar flares have been proposed to work in different flares, leading to different types of QPPs. Several different physical processes may be responsible for the generation of QPPs, which can be summarized into the following three categories: (1) oscillatory (including MHD oscillations, QPPs triggered periodically by external waves, dispersive wave trains, etc.) \citep[e.g.,][]{2006A&A...452..343N}; (2) self-oscillatory (including periodic spontaneous reconnection, coalescence of two magnetic flux tubes, etc.); and (3) auto wave processes. \lf{A} schematic illustration of the main models interpreting solar flare QPPs can be found in \citet{2020STP.....6a...3K}.  \ya{Many QPPs are identified based on the light curve of a region. Flare-related multiple periodic pulsations can be detected using multiple wavelengths and the generation mechanism is complicated \citep[]{2021ApJ...921..179L}. }\lf{Observational knowledge of the} \ya{spatially-resolved} source region of flare QPPs is needed.\\
%by the capabilities of instruments. Thus, corresponding physical properties deserve further study with high-resolution observations and some unique observational wavelengths. \\

In recent years, some studies reported \lf{a} relationship between the three-minute oscillation and energetic events \citep[e.g.,][]{2007ApJ...670L.147K, 2009A&A...505..791S, 2017ApJ...848L...8M, 2021MNRAS.503.2444M}.  \citet{2009A&A...505..791S} provided observational evidence of the leakage of three-minute oscillations into the corona along the coronal loops, and 
\lf{proposed that these are involved in triggering the QPPs in the energy release.} 
\citet{2021MNRAS.503.2444M} studied the variation of \lf{the location and period of} chromospheric oscillations during a solar flare, and by comparing the pre-flare and post-impulsive behaviour they provided evidence \lf{that} the change in the magnetic environment caused by a solar flare can affect the \lf{oscillations.}
%solar atmosphere, even the whole active region.
\citet{2017ApJ...848L...8M} studied the three-minute oscillation in Ly$\alpha$ and Ly$C$ emission during an X-class flare \lf{finding evidence} that compressible waves with a period around the acoustic cutoff are created when the chromosphere is impulsively disturbed. \smm{In order to understand the nature of the three-minute oscillation above the sunspot during flares, we require further study using high-resolution observations and magnetohydrodynamics (MHD) numerical simulation.}\\

\lf{In this paper, we analyse a strong intensity oscillation observed at high resolution in a flare ribbon at an umbral-penumbral boundary.} The flare event, SOL2012-07-05T21:42, was an M1.8 class X-ray flare, other aspects of which have been studied from three perspectives. These are 1) the sunspot dynamics during the flare \citep[Part~I,][]{2016ApJ...833..250W}, 2) the origin and destination of multiple hot channels \citep[Part~II,][]{2018ApJ...859..148W}, and 3) the EUV late phase \citep[Part~III,][]{2020ApJ...905..126W}. We mentioned \lf{an apparent `back and forth' oscillation of the flare ribbon's position} in the study of Part~I, \lf{while in Part~III, we highlighted the presence of strong downflows in the region, including its influence on the loop which we study in this work.} \\

Here we study and discuss the enhanced three-minute oscillation in multiple passbands at the sunspot's umbral and penumbral boundary, where the footpoint of the late phase loops and the flare ribbon arrives.  %and try to reveal 
%\smm{We examine} the relationship between the three-minute oscillation in a sunspot and the energy-release process of a flare.
%LF I took this out because I don't think that we really have anything to say about the energy-release process!
We perform a joint analysis of data from ground-based and space-based instruments together, \lf{to characterise the oscillation properties before and after the flare, at wavelengths corresponding to emission from the deep chromosphere to coronal temperatures.} In this way, we find the connection between waves in the chromosphere and the corona through direct comparison of the oscillation variations by studying their spatial and temporal properties both visually and using cross-correlation.  Section~\ref{sec2:overview} \smm{provides an} overview of the active region, and we present our analysis and results in section~\ref{sec3:results}. The conclusion and discussion are presented in section~\ref{sec4:conclusion}. \\

%%%%%%%%%%%%%%%%%%%%%%%%%%%%%%%%%%%%%%%%%%%%%%%%%
\section{Data and overview of the active region} 
\label{sec2:overview}
%%%%%%%%%%%%%%%%%%%%%%%%%%%%%%%%%%%%%%%%%%%%%%%%%

\begin{figure*}
\begin{center}
% trim=left bottom right top
\includegraphics[ width=0.9\textwidth]{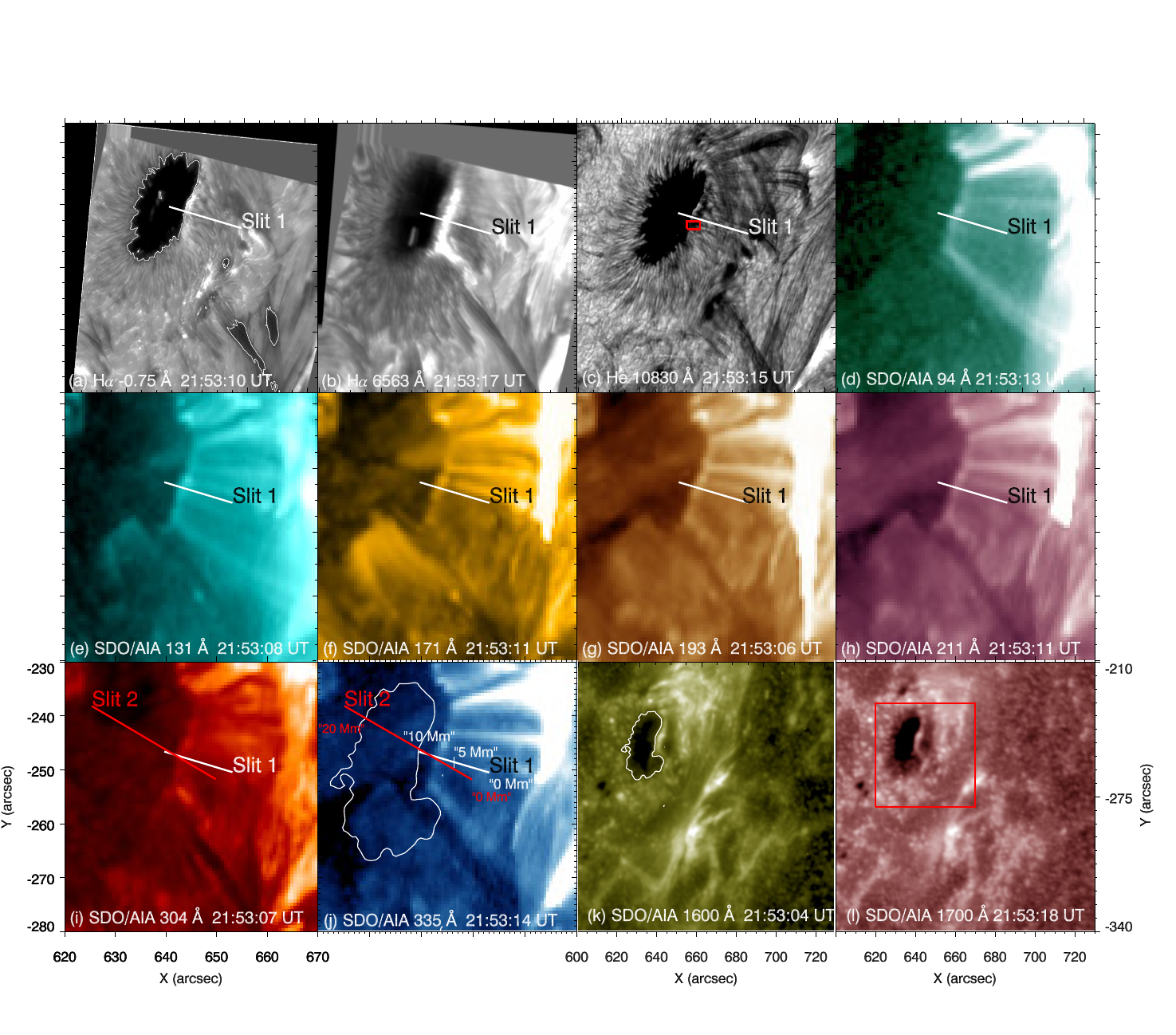}
\caption{\smm{Panels (a$-$c): The near-simultaneous images of the sunspot and the flare emission are shown at multiple wavelengths, H$\alpha$ blue wing -0.75~\AA, H$\alpha$ 6563~\AA\ and He~\textsc{i} 10830~\AA\ from BBSO/GST. The white contour in panel (a) represents the contour of the sunspot's umbra (for DN~s$^{-1}$ = 1000). The red box in panel (c) was used to create the intensity profiles in Fig.~\ref{fig:fig2}. Panels (d$-$l): EUV and UV images obtained from the AIA passbands $-$ 94, 131, 171, 193, 211, 304, 335, 1600 and 1700~\AA\ respectively. The red boxed region in panel (l) represents the FOV for all AIA EUV images shown in panels (d-j). The white (red) solid lines, slits 1 and 2 indicate artificial slits that were used to create the time-distance plots shown in Figs.~\ref{fig:fig4} (and~\ref{fig:fig3}). Slits 1 and 2 lengths are 10 and 20~Mm, respectively. The contour in panel (k) represents the outline of the sunspot (including umbra and penumbra for the intensity of the AIA 1600~{\AA} = 75 DN~s$^{-1}$) and we plotted the same contours in panel (j). Please note, panels (k) and (l) have a different scale from the others.} \label{fig:fig1}}
\end{center}
\end{figure*}

The observational data in this paper are the same as those previously introduced in \cite{2016ApJ...833..250W, 2018ApJ...859..148W, 2020ApJ...905..126W}. High-resolution \smm{ground-based images were obtained in} He~\textsc{i} 10830~\AA\ narrowband (bandpass: 0.5~\AA) \smm{from} the Big Bear Solar Observatory \smm{(BBSO)} with the 1.6-meter aperture Goode Solar Telescope \smm{(GST)} \citep[]{2010AN....331..620G, 2012SPIE.8444E..03G}. \smm{For the He~\textsc{i} 10830~\AA\ images,} the pixel size is 0.$\arcsec$0875, the cadence is around 10~s, and the field-of-view (FOV) is about 90$\arcsec$$\times$90$\arcsec$. \smm{In addition,} high-resolution H$\alpha$ images at the line center, \smm{6563~{\AA}} and \smm{H$\alpha$} -0.75~\AA\ blue wing were also taken from BBSO/GST. For the H$\alpha$, the pixel size is 0.$\arcsec$056. The cadence is about 34~s for the H$\alpha$ line center and 10~s for the H$\alpha$ blue wing. Long-duration good seeing conditions at BBSO and the high-order adaptive optics (AO) system are beneficial \lf{for obtaining} consecutive diffraction-limited images lasting for hours. The seeing conditions were good on July~5, 2012, and the NOAA Active Region (AR) \smm{\#}11515 (S17W37)\footnote{https://www.solarmonitor.org/?date=20120705} was selected as the observation target. \smm{The active region consisted of a sunspot that had a complex magnetic configuration of $\beta\gamma\delta$\ya{. The FOV of GST contains the western part of the AR, which is} dominated by a positive magnetic field and surrounded by satellite sunspots with scattered negative fields. The Geostationary Operational Environment Satellite (GOES-15) recorded an M1.8 class X-ray flare originating from this active region at 21:37~UT. The flare peaked at 21:42~UT and ended at 22:30~UT. Based on the EUV-integrated light curves shown in Fig.~2 of \cite{2020ApJ...905..126W}, we identify the three phases of the flare: the pre-flare phase (between 21:00 and 21:36~UT), the impulsive phase (between 21:37 and 21:45~UT), and the EUV late phase (between 21:55 and 22:30~UT). \lf{The EUV late phase is a second gradual-phase peak in the warm EUV irradiance, such as Fe~\textsc{xvi} 335~{\AA} ($\sim$3~MK) discovered by \cite{2011ApJ...739...59W} using the EUV Variability Experiment (EVE) instrument \citep{2012SoPh..275..115W}  on board the Solar Dynamics Observatory (SDO)}. The main sunspot associated with the AR \#11515 is shown in Fig.~\ref{fig:fig1}, in H$\alpha$ blue wing -0.75~{\AA} (panel~a), H$\alpha$ 6563~\AA\ (panel~b) and He~\textsc{i} 10830~\AA\ (panel~c)  passbands at 21:53~UT. }

%%%%%%%%%%%%%%%%%%%%%%%%%%%%%%%%%%%%%%%%%%%%%%%%
% Figure 1
%%%%%%%%%%%%%%%%%%%%%%%%%%%%%%%%%%%%%%%%%%%%%%%%
%%%%%%%%%%%%%%%%%%%%%%%%%%%%%%%%%%%%%%%%%%%%%%%%%%%%%%%

%%We used a series of SDO/AIA data corresponding to the ground-based telescope observations. For the EUV/UV images observed by the AIA \citep{2012SoPh..275...17L} on board the SDO,
\smm{We studied this flare using the UV/EUV images obtained from the \textit{Atmospheric Imaging Assembly} instrument \cite[AIA; ][]{2012SoPh..275...17L} on board SDO \citep{2012SoPh..275....3P} during the time interval when the ground-based data were available.} \smm{AIA captures full-disk images of the solar disk (4096~$\times$~4096~pixels) every 12~s in 7 EUV and every 24~s in 2 UV passbands with a spatial resolution of 1.$\arcsec$2 (pixel size of 0.$\arcsec$6)}. Seven EUV passbands \citep{2010A&A...521A..21O, 2013A&A...558A..73D} are 94~\AA\ (Fe~\textsc{xviii}, \smm{log~\textit{T}~[K] = }6.8), 131~\AA\ (Fe~\textsc{viii, xxi}, \smm{log~\textit{T}~[K] =} 5.6, 7.0), 171~\AA\ (Fe~\textsc{ix}, \smm{log~\textit{T}~[K] = }5.8), 193~\AA\ (Fe~\textsc{xii, xxiv}, \smm{log~\textit{T}~[K] = }6.2, 7.3), 211~\AA\ (Fe~\textsc{xiv}, \smm{log~\textit{T}~[K] = }6.3), 304~\AA\ (He~\textsc{ii}, \smm{log~\textit{T}~[K] = }4.7), and 335~\AA\ (Fe~\textsc{xvi}, \smm{log~\textit{T}~[K] = }6.4). They are sensitive to the corona and transition region, except for 304~\AA, which is sensitive to the chromosphere. In addition, UV images with 24~s cadence were obtained from \smm{AIA} 1600 \AA\ and 1700 \AA\ passbands. \lf{In the quiet Sun these channels are sensitive to photosphere, chromosphere and transition region. The flare excess emission in the 1600 \AA\ channel is dominated by C~\textsc{iv} (\smm{log~\textit{T} [K] = }5.0) and Si continuum, and in the 1700 \AA\ channel by C~\textsc{i} (\mbox{log~\textit{T} [K] = }4.2) and  He~\textsc{II}} \citep{2019ApJ...870..114S}.

\smm{The near-simultaneous images in UV and EUV passbands are shown at 21:53~UT in Fig.~\ref{fig:fig1} (panels d$-$l). Note that the FOV shown in AIA 1600 and 1700~{\AA} images is bigger than the FOV shown in other panels in this figure. The red boxed region highlighted in panel (l) indicates the FOV shown in H$\alpha$, He~\textsc{i} and all EUV passbands. The sunspot is highlighted in panel (k) with white contours and the same contours are overplotted in the AIA~335~{\AA} image in panel (j). Slit 1 (with length 10~Mm) is plotted in these panels which crossed the umbral-penumbral boundary of the sunspot. The region captured in this slit was further used to identify the oscillations in EUV passbands (see section~\ref{subsec:oscillation_phenomenon} and Fig.~\ref{fig:fig4}). \text{Slit 2} of length 20~Mm is shown in red colour in panels (i$-$j) which was crossing both umbral-penumbral boundaries of the sunspot where we identified the running penumbral waves (see Fig.~\ref{fig:fig3}).}

%The AIA temperature coverage is thus from 1.6$\times$10$^{4}$ to 2$\times$10$^{7}$ K. 

%The hmi.M$_{-}$45s data series provide us full-disk magnetograms along the line of sight with a cadence of 45 s. 

As \lf{previously} analyzed \citep{2016ApJ...833..250W, 2018ApJ...859..148W, 2020ApJ...905..126W}, a large-scale filament rooted in the sunspot erupted and was accompanied \smm{by} a \smm{coronal mass ejection} (CME). Three sets of loops, mutually confirmed by AIA imaging observations and the magnetic topology from \smm{nonlinear force-free field} (NLFFF) extrapolations, include a set of post-flare loops and two sets of late-phase loops \ya{(see Figs.~3 and 4 in} \citet{2020ApJ...905..126W}).  \ya{A main flare ribbon and a secondary flare ribbon are observed at around 21:42 UT (see Fig.~1 in \cite{2020ApJ...905..126W}).} The EUV late phase dominates the gradual phase observed in EUV passbands, especially the \ya{335 \AA}, which \lf{show} second peaks after the impulsive phase. During the flare, one of the flare ribbons swept into the sunspots and oscillated. Based on our previous studies, \lf{we identify} three sets of loops in the region. Two of them are the post-flare loop and EUV late-phase loop, respectively. The footpoints of both loop sets are at the location of oscillation. The late-phase loop connected the flaring site and a remote plage region, which is located at an asymmetric quadrupole magnetic field configuration. During the EUV late phase, downflowing plasma is observed, which collided with the materials in the low-lying atmosphere \ya{(see Figs.~7 and 8, and the corresponding animation in \cite{2020ApJ...905..126W}).} In Part~III \citep{2020ApJ...905..126W}, we proposed that this could be an additional heating mechanism of the EUV late-phase loops.  \\

%%%%%%%%%%%%%%%%%%%%%%%%%%%%%%%%%
\section{Analysis and Results} 
\label{sec3:results}
%%%%%%%%%%%%%%%%%%%%%%%%%%%%%%%%%

%%%%%%%%%%%%%%%%%%%%%%%%%%%%%%%%%
%\begin{figure}
%\includegraphics{plot_ir1083_lightcurve_eps_1.pdf}
%\caption{The time profile of the intensity integrated over the red box in Fig. 1(c) at He~\textsc{i} 10830~\AA. The red curve overlaid on the profile is the slow component for the time profile, the three arrows indicate the two precursors and the flare impulsive phase where the absorption reaches its valley. Panel (b) represent the fast component for the intensity obtained from the time profile in panel (a); Panel (c) shows the result of wavelet analysis.
%\label{fig:fig2}}
%\end{figure}
%%%%%%%%%%%%%%%%%%%%%%%%%%%%%%%%%

%%%%%%%%%%%%%%%%%%%%%%%%%%%%%%%%%%%%%%%%%%%%%%
\subsection{The Oscillation Phenomenon}
\label{subsec:oscillation_phenomenon}
%%%%%%%%%%%%%%%%%%%%%%%%%%%%%%%%%%%%%%%%%%%%%%

%What we need to clarify is that the oscillation of the flare ribbon refers to the oscillation of the footpoint of the late-phase loop. Because the location of the footpoint of the late-phase loops overlaps visually with the position of one ribbon of the two-ribbon flare. The difference is that they show up at different times. So, there are two kinds of names in this paper. They are the oscillation of the flare ribbon and the oscillation of the footpoint of the late-phase loop.

\lf{During the impulsive phase of the flare} the eastern branch of the two ribbons \lf{advanced rapidly} into the sunspot and oscillated \lf{in position} at the boundary between the umbra and penumbra. Fig.~\ref{fig:fig1} shows the ribbon position after the impulsive phase. The flare ribbon shows emission in all passbands except for the He~\textsc{i} 10830~\AA\ passband, where it is in absorption and shows up dark. 

\begin{figure*}
\begin{center}
% trim=left bottom right top
\includegraphics[trim=1.3cm 0cm 0.5cm 0.9cm, width=0.48\textwidth]{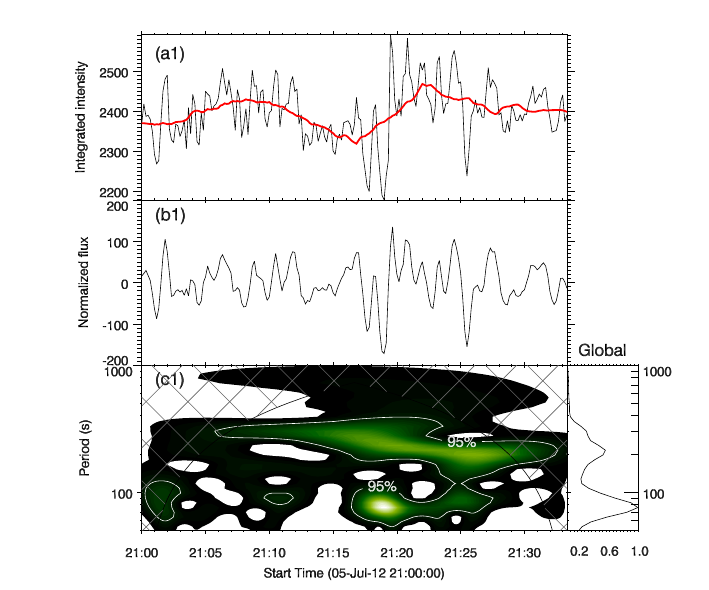}
\includegraphics[trim=0.6cm 0cm 1.2cm 0.9cm, width=0.48\textwidth]{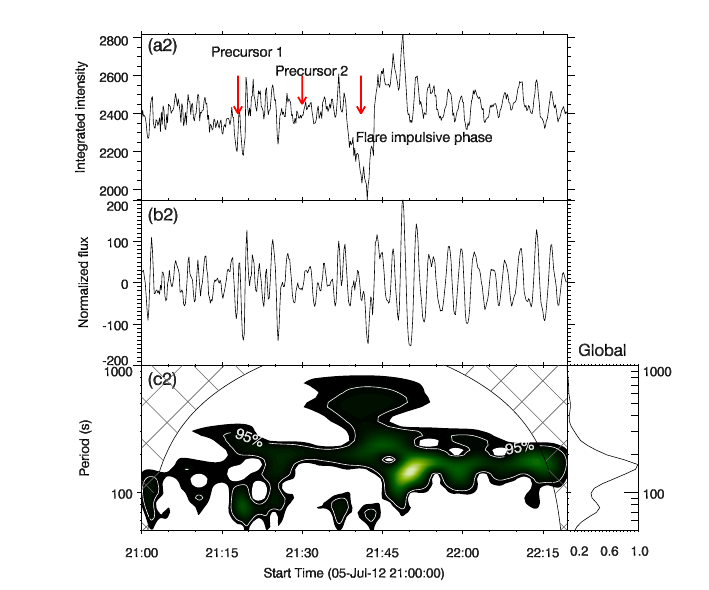}
\caption{\smm{He~\textsc{i} 10830~{\AA} time profiles of the intensity integrated over the red boxed region shown in Fig.~\ref{fig:fig1} (panel~c). They are plotted for} \lf{the pre-flare phase} from 21:00 to 21:33~UT (panels a1-c1) and \lf{over the whole event} from 21:00 to 22:20~UT (panels a2-c2). The red curve overlaid on the profile \smm{shown in panel~(a1)} is the slow component for the time profile. The three arrows \smm{in panel~(a2)} indicate \lf{times of the} two precursors \smm{(at 21:20 and 21:30~UT)} and the flare impulsive phase \smm{(at 21:42~UT)} where the absorption reaches its trough. Panels (b1) and (b2) represent the fast component for the intensity obtained from the time profile in panels (a1) and (a2); Panels (c1) and (c2) show the wavelet analysis of the power spectrum using the Morlet wavelet. \ya{The power is normalized. The white contour indicates a 95\% confidence level and the bright region inside the white contour indicates a region with greater than 95\% confidence level for a white-noise process. Cross-hatched regions indicate the cone of influence, where the edge effect becomes important. The peak at around 80 s (panel c1) is produced by a microflare  at $\sim$21:20~UT, which is regarded as the first precursor. The global wavelet power spectra are also shown.}
\label{fig:fig2}}
\end{center}
\end{figure*}

We first \lf{remarked on} the oscillation of the ribbon position in the He~\textsc{i} 10830~\AA\ passband \lf{in \cite{2016ApJ...833..250W}}. \lf{To further investigate the oscillatory behaviour}, we took a small region shown by the red box in Fig.~\ref{fig:fig1}, panel~c \smm{and obtained He~\textsc{i} time profile of the integrated intensity. The results are shown in Fig.~\ref{fig:fig2}}. The detrended intensity \lf{in this region of the ribbon} presents significant oscillations \lf{which can be characterized using a} Morlet wavelet analysis \citep{1998BAMS...79...61T}. \ya{Panels (a1) and (a2)  present the raw time series integrated over the red box shown in Fig.~\ref{fig:fig1}, panel~c. The overlaid red curve in panel (a1) presents the slow-varying trend, and the detrended time series is shown in panels (b1) and (b2). Here, we used smoothing boxcars with widths of 5~minutes and 3~minutes} for the raw light curves in panels (a1) and (a2), respectively, to remove the large-scale trend. The low-frequency components are removed. Detrending is often utilized in time series before analysis of their period or frequency in order to strengthen the periodic signals in raw time series \citep[]{2016ApJ...825..110A}. We performed the wavelet analysis for the detrended lightcurves (panels (b1) and (b2)). The parameter $\omega$$_{0}$ for the Morlet wavelet is set to 6 by default. We prefer to use a small (the default) value in order to have a good time resolution, because of fast variation around the flare. The wavelet power spectra show a dominant period of 3-4 minutes which is above the 95\% significance level. \lf{The times of two microflares (which are regarded as two precursors), and the flare impulsive phase, respectively 21:20, 21:30, and 21:42~UT are indicated with red arrows on the lightcurve. The first precursor and the impulsive phase correspond to dips in the light curve (Fig.~\ref{fig:fig2}, panel a2). Compared with the pre-flare stage before 21:30~UT (Fig.~\ref{fig:fig2}, panels a1-c1}), the three-minute oscillation is enhanced from 21:50 to 22:15~UT (Fig.~\ref{fig:fig2}, panel~c2), which is right in the EUV late phase of the flare. \lf{Additionally}, a frequency shift occurs before and after the impulsive phase of the flare.  We note \lf{first of all} that the \lf{dominant} period before 21:15~UT is about 5~minutes because the He~\textsc{i} 10830~\AA\ images at this time \lf{likely includes also} the photospheric \lf{emission} from the quiet Sun. \lf{Then, between 21:20 and 21:33~UT (after the first precursor) the dominant period is a chromospheric period of about 200~s. This} decreased to 180~s between 21:50-22:15~UT. The interpretation will be \lf{described} in detail in the following Sections.\\

%%%%%%%%%%%%%%%%%%%%%%%%%%%%%%%%%%%%%%%%%%%%%%%%%%%%%%% 
% Figure 2
%%%%%%%%%%%%%%%%%%%%%%%%%%%%%%%%%%%%%%%%%%%%%%%%%%%%%%% 
%%%%%%%%%%%%%%%%%%%%%%%%%%%%%%%%%%%%%%%%%%%%%%%%%%%%%%%

\begin{figure}
\begin{center}
% trim=left bottom right top
\includegraphics[trim=0cm 0.6cm 0cm 1.1cm, width=0.55\textwidth]{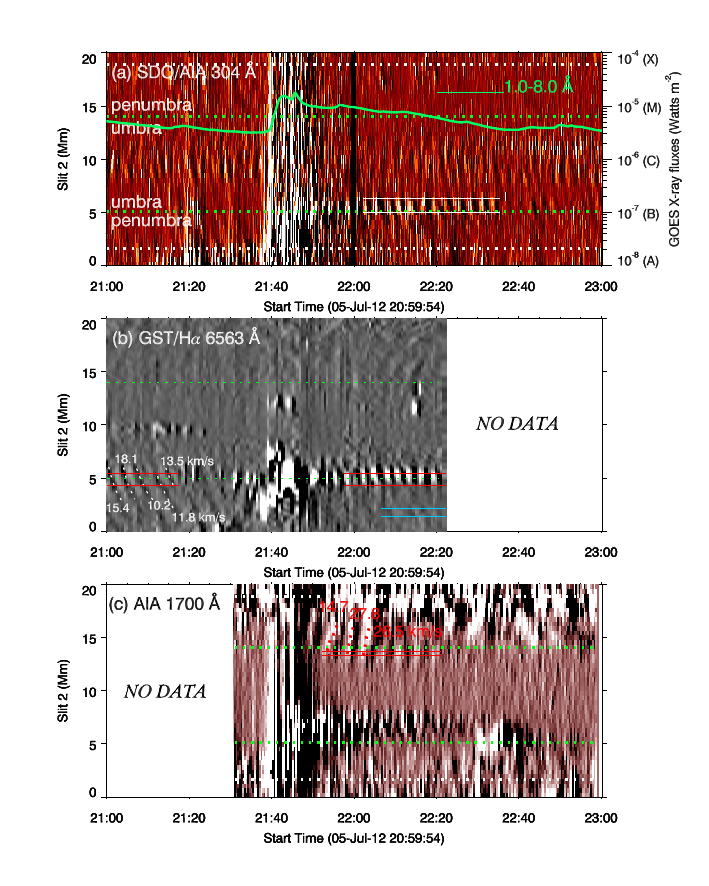}
\caption{\smm{The time-distance plots were obtained along \lf{slit 2} (the red slit is shown in panel (i) of Fig.~\ref{fig:fig1}) by stacking running-difference images in the AIA 304 (panel~a), H$\alpha$ 6563~\AA\ (panel~b), and  AIA 1700~{\AA} (panel~c) channels}. The green curve in panel~(a) indicates the soft X-ray \smm{fluxes obtained in 1-8~\AA\ channel of the GOES-15}. The green dotted lines indicate the boundary between the sunspot's umbra and penumbra and the white dotted lines represent the \smm{outer} boundary of the penumbra. \smm{The region of enhanced oscillation observed between 22:00 and 22:35~UT is shown by two white (red) parallel lines in AIA 304~{\AA} (H$\alpha$) image. The running penumbral wave (observed as grey and black stripes between two red horizontal lines during 21:00 and 21:20~UT) propagates at a speed of 10.2$-$18.1~{$\rm km\,s^{-1}$} before the flare onset in H$\alpha$ whereas the penumbral waves were observed on the other side of the sunspot in AIA 1700~{\AA} (observed as black and white stripes at $\sim$13~Mm shown between two red horizontal lines during 21:50 and 22:20~UT) and propagated with speed between 14.7 and 27.8~{$\rm km\,s^{-1}$}.} Between each two sets of parallel lines in panels (a)-(c), we made the integrated intensity. The profiles and corresponding wavelet analysis are made as shown in Fig.~\ref{fig:fig9}.\label{fig:fig3}}
\end{center}
\end{figure}

To compare the \lf{intensity} oscillation in the sunspot's umbra and penumbra, and the oscillation before and after the flare, we made \lf{a slice} across the sunspot (\smm{at} slit 2 \smm{location shown} in red colour in  \smm{panels (i) and (j)} of Fig.~\ref{fig:fig1}) with a length of about 20~Mm and used this to create time-distance plots in \smm{H$\alpha$ 6563~{\AA} and AIA 1700~{\AA}, 304~{\AA} passbands}, which are shown in Fig.~\ref{fig:fig3}.  \ya{The GST/H$\alpha$ data is unavailable after 22:22 UT and the SDO/AIA 1700 \AA\ data is unavailable before 21:30 UT. For AIA 1700 \AA\ data,  we cannot study the RPW before the flare due to the lack of data. However, after the flare we can study the properties of the RPW that is clearly visible in 1700 \AA\ on the other side of the sunspot.}

%\lf{[We should mention the missing data in this figure, and how it affects our choice of `before' and `after' times.]} 

\smm{In Fig.~\ref{fig:fig3} panel (a), we show the time-distance plot in} \lf{the} AIA 304~\AA\ passband overlaid (shown as green curve) by \smm{the X-ray fluxes observed by GOES-15 in the 1-8~\AA\ channel}. \smm{The extent of the umbral and penumbral boundaries are indicated by dotted green and white horizontal lines respectively in all images.} We clearly observe the umbral wave \smm{(between 8 and 12~Mm)} in the sunspot's umbra, propagating from the center of the umbra towards the penumbra, \lf{with a characteristic pattern} like a stack of bowls. The edge of the bowl extends to the penumbra. \smm{In AIA 304~{\AA} passband, we did not observe the running penumbral waves between 21:00 and 21:40~UT. During the EUV late phase}, we find that there is enhanced oscillation at the boundary between the sunspot's umbra and penumbra from $\sim$22:00 to $\sim$22:35~UT \smm{(the region is shown between two solid white lines)}. This is consistent with the enhanced signal of three-minute oscillation on the integrated light curve of the He~\textsc{i} 10830~{\AA} \smm{(see Fig.~\ref{fig:fig2})

In Fig.~\ref{fig:fig3} panel (b), we show the H$\alpha$~6563~{\AA} time-distance plot} and the running penumbral wave \smm{(observed as black and grey stripes between two red horizontal lines running from 8 to 4~Mm during 21:00 and 21:20~UT)} propagates at a speed of 10.2$-$18.1~{$\rm km\,s^{-1}$} before the flare onset. \ya{The enhanced oscillation (shown by two red horizontal lines during 22:00 and 22:20~UT) and the RPW (shown as two blue lines during 22:07 and 22:22~UT) co-exist during the late phase of the flare. Fig.~\ref{fig12} in the appendix shows the co-existence of the two features.} While a part of the running penumbral waves converts into oscillation \smm{during the EUV late phase}, the other part of the wave \lf{maintains} the same pattern as that before the flare. 

In Fig.~\ref{fig:fig3} panel (c), we showed the time-distance plot in 1700~\AA. We observed the penumbral wave propagating on the other side of the sunspot \smm{(observed at $\sim$13~Mm between 21:50 and 22:20~UT)}. The \lf{propagation} speeds are from 14.7 to 27.8~{$\rm km\,s^{-1}$} \lf{found} by linear fitting \lf{of} the pattern of the motion. We also find an enhanced fluctuation at the \lf{opposite} boundary between the sunspot's umbra and penumbra during the same period where the signal is seen in the \ya{1700}~\AA\ passband \ya{(see black and white stripes around 5~Mm in panel~(c) of Fig.~\ref{fig:fig3} during 21:50 and 22:30~UT)}. However, the observed fluctuation seems to be irregular.  \\

%%%%%%%%%%%%%%%%%%%%%%%%%%%%%%%%%%%%%%%%%%%%%%%%
% Figure 3
%%%%%%%%%%%%%%%%%%%%%%%%%%%%%%%%%%%%%%%%%%%%%%%%
%%%%%%%%%%%%%%%%%%%%%%%%%%%%%%

\begin{figure*}
\begin{center}
% trim=left bottom right top
\includegraphics[trim=0.5cm 0.1cm 0cm 0.5cm, width=0.95\textwidth]{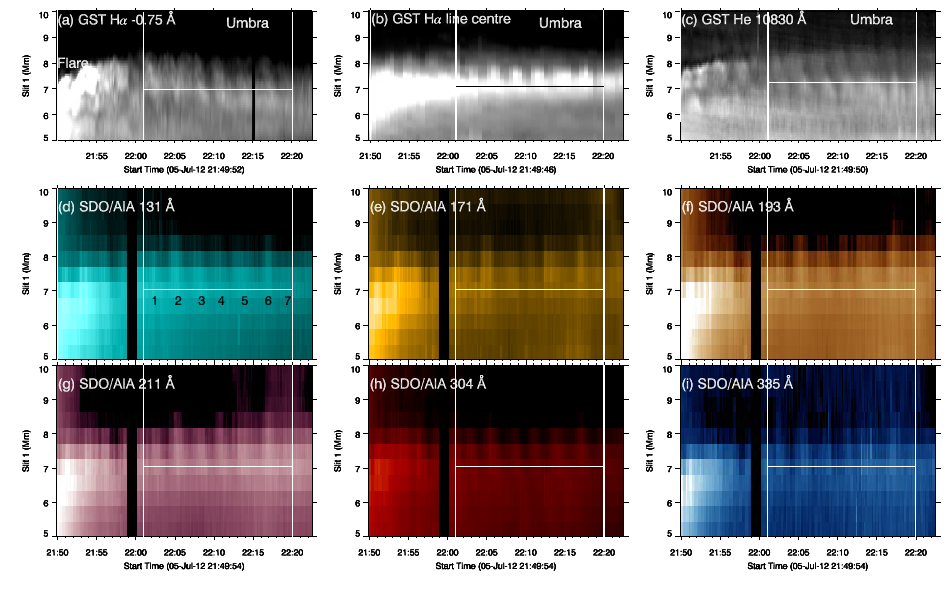}
\caption{\smm{The time-distance plots obtained along slit 1 (shown as a white slit in panels~a$-$j in  Fig.~\ref{fig:fig1}) by stacking intensity images in} H$\alpha$ -0.75, H$\alpha$ 6563~{\AA} line center, He~\textsc{i} 10830, AIA 131, 171, 193, 211, 304 and 335~\AA\ wavelength channels. \smm{The dark region above 8.5~Mm \lf{corresponds to} the sunspot's umbra and the bright region is flare emission (see panel (a)).} The oscillation contains at least seven peaks from 22:00~UT to 22:20~UT \smm{and they are labelled in panel (d)}. The intensity variation profiles are obtained along the white horizontal lines and are shown in Fig.~\ref{fig:fig5}. The white vertical lines indicate the time range of the intensity variation shown in Fig.~\ref{fig:fig5}. \label{fig:fig4}}
\end{center}
\end{figure*}

\smm{We made slices along the direction of the oscillation to obtain the time-distance plots in Fig.~\ref{fig:fig4} i.e. at the position of slit 1 that is indicated by white lines in Fig.~\ref{fig:fig1} (panels~a$-$j)}. The slices are made almost perpendicular to the flare ribbon. \lf{In these} time-distance plots (see Fig.~\ref{fig:fig4}), the dark region above 8.5~Mm \lf{corresponds to} the sunspot's umbra and \smm{the bright region is emission from the late phase of the flare}. The oscillation lasted for more than one and a half hours, which covers the range of the EUV late phase. \smm{We observed at least seven \lf{oscillation pulses} and they are indicated by numbers (1$-$7) in the 131~\AA\ passband (see Fig.~\ref{fig:fig4}, panel (d)). By examining the imaging observation in H$\alpha$ 6563~\AA\ and EUV passbands, we confirm that the} intensity oscillation \lf{is unambiguously present} in both chromospheric and coronal passbands. \\

%%%%%%%%%%%%%%%%%%%%%%%%%%%%%%%%%%%%%%%%%%%%%%%%
% Figure 4
%%%%%%%%%%%%%%%%%%%%%%%%%%%%%%%%%%%%%%%%%%%%%%%%
%%%%%%%%%%%%%%%%%%%%%%%%%%%%%%%%%%%%%%%%%%%%%%%%%%%%%%%

\begin{figure*}
\begin{center}
% trim=left bottom right top
\includegraphics[trim=1.3cm 0cm 0.5cm 0.9cm, width=0.5\textwidth]{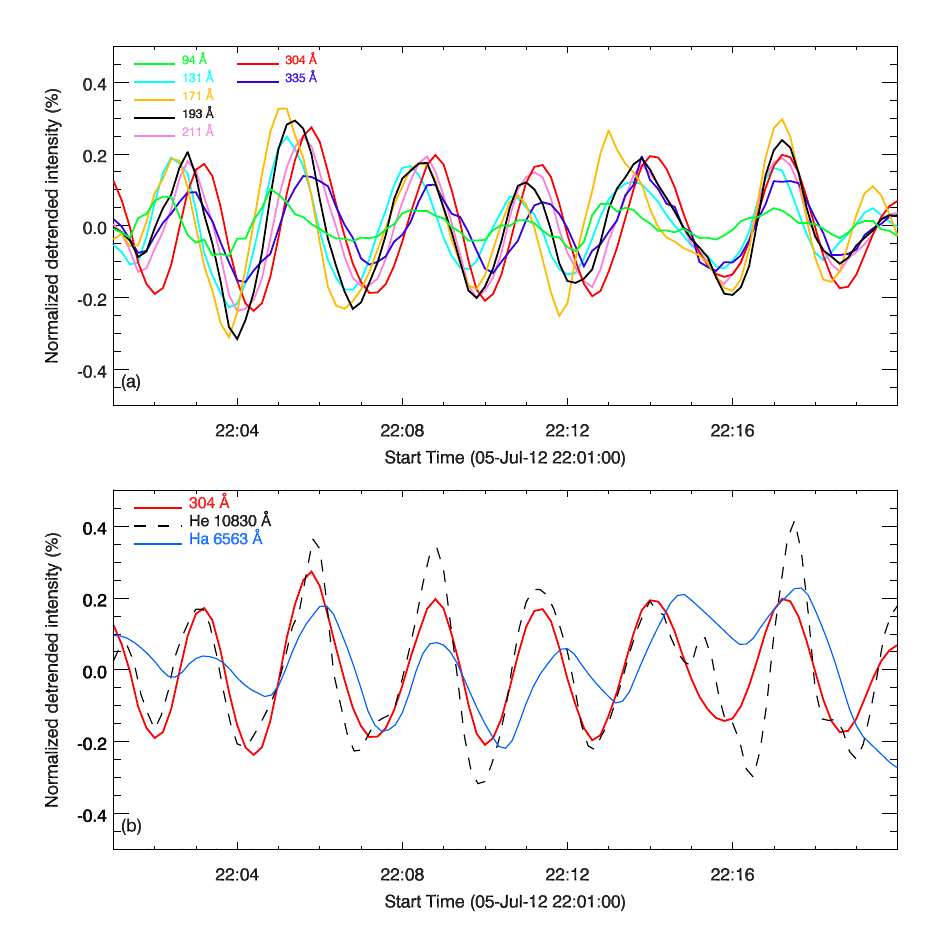}
\caption{\smm{The intensity variation profiles for all AIA EUV (top panel), H$\alpha$ and He~\textsc{i} passbands (bottom panel) obtained along the white horizontal lines indicated in Fig.~\ref{fig:fig4}. } \label{fig:fig5}}
\end{center}
\end{figure*}

\smm{We obtained the intensity variation profiles for all AIA EUV, H$\alpha$ and He~\textsc{i} passbands during 22:00 and 22:20~UT along the white horizontal lines indicated in Fig.~\ref{fig:fig4} and the results are shown in Fig.~\ref{fig:fig5}.} There are obvious time lags (phase differences) between each passband. \smm{We have taken the AIA 94~{\AA} passband as a reference and obtained time lags in all AIA passbands using cross-correlation and the results are presented in Table~\ref{Table1:time_lag_info}. In addition, we have taken the AIA 304~{\AA} passband as a reference and obtained time lags for H$\alpha$ and He~\textsc{i} passbands and the results are summarized in Table~\ref{Table2:time_lag_info}.} We found that the peak of the oscillation \lf{in the intensities} occurs first in high-temperature \lf{EUV} passbands like 94 and 131~\AA, and then in some \lf{cooler EUV} passbands. Finally, it occurs in the chromospheric  H$\alpha$ 6563~\AA\ line. However, the \lf{oscillatory} behaviour of He~\textsc{i} 10830~\AA\ line is quite consistent with the AIA 304~\AA\ passband \smm{(dominated by He~\textsc{ii} line)}, rather than the chromospheric H$\alpha$ 6563~\AA\ line because the formation of He~\textsc{i} 10830~\AA\ is complicated. He~\textsc{ii} 304~\AA\ is singly ionized helium
%The formation of He~\textsc{i} 10830~\AA\ is affected by recombination following photoionization and collisional ionization \citep[e.g.][]{Kerr2021} and collisional exitation . This could be the reason why the intensity evolution of He~\textsc{i}~10830~\AA\ is similar to the He~\textsc{ii} 304~\AA. \lf{[LF: I don't understand this part. Are we saying that photoionisation and collisional ionisation directly produce the He II ions and also affect the production of the 10830? I think this needs to be explained a little more. Note, the paper of Kerr et al. that I added says that collisional ionisation by non-thermal flare electrons is more important in a flare, but I don't think that this is necessarily true during the gradual phase. We probably need an additional reference on He 10830 formation here. ]}
\ya{and He~\textsc{i} 10830~{\AA} corresponds to the transition between 1s2s $^{3}$S and 1s2s $^{3}$P of the helium triplet. For the formation mechanism of the helium triplet, the photoionization then recombination mechanism (PRM), collisional ionization then recombination mechanism (CRM), and collisional excitation mechanism (CM) are widely reported \citep[e.g.][]{1997ApJ...489..375A, 2021ApJ...912..153K}. For the PRM mechanism \citep[e.g.][]{1975ApJ...199L..63Z, 1994IAUS..154...35A, 2008ApJ...677..742C, 2016A&A...594A.104L}, coronal photons at wavelengths which shorter than 504~{\AA}, penetrate into the chromosphere and photoionize helium atoms. These atoms then recombine free electrons to the excited levels of a triplet of Helium atom. While, for the CRM mechanism \citep[e.g.][]{2005A&A...432..699D},  the nonthermal electrons can collisionally ionize the helium atom during a flare. The two processes can lead to the overpopulation of ionized helium and excited levels of triplet helium. The non-thermal collisional ionization plays an important role in the impulsive phase of a flare. However, once the temperature has increased during the flare, the thermal effect becomes important. The thermal collisional ionization and recombination can affect the population of triplet helium \citep[e.g.][]{2021ApJ...912..153K} .}\\

%For all the passbands, we fitted the oscillations by using the sine function as follows. \\
%$F = A_{0}*sin(A_{1}*x+A_{2})+A_{3}$\\
%\[EM_{c}=\int_{T_{min}}^{T_{max}} DEM(T)\,dT=\int_{}^{}n_e^2dh\]
%Based on the fitting results, the period of oscillation is about three minutes. The amplitude is estimated to be a few hundred kilometers. The oscillation keeps the same both in the chromosphere and in the corona. \textbf{We need to obtain the period, amplitude and phase of the oscillation, then compare the results among all the passbands in detail.} \\

%600 km for the EUV wavelength.

 %While the oscillation observed in AIA 304~\AA\ also presents a 3-minute oscillation. Based on the observation from Ha 6563~\AA, the period before and after the flare impulsive phase presents a slight difference, which will be discussed in the following.  \\

%As shown in Fig. 6, we suggest that the oscillation occurs on the footpoint of the EUV late-phase loop which connects the sunspot and the bipolar field. The 3-minute oscillation can be observed from the chromosphere to the corona.  \\

%%%%%%%%%%%%%%%%%%%%%%%%%%%%%%%%%%%%%%%%%%%%%%%%%%%%%%% 
% Figure 5
%%%%%%%%%%%%%%%%%%%%%%%%%%%%%%%%%%%%%%%%%%%%%%%%%%%%%%% 
%%%%%%%%%%%%%%%%%%%%%%%%%%%%%%%%%%%%%%%%%%%%%%%%%%%%%%%

%%%%%%%%%%%%%%%%%%%%%%%%%%%%%%%%%%%%
%%% Table 1
%%%%%%%%%%%%%%%%%%%%%%%%%%%%%%%%
\begin{table}[htbp]
\centering
\caption{Time lags between each \smm{AIA} EUV passband}
\setlength{\tabcolsep}{0.9mm}{ 
\begin{tabular}{c|ccccccc}
\hline \hline
AIA &94 \AA &131 \AA & 171 \AA &193 \AA & 211 \AA & 335 \AA& 304 \AA  \\
% &   &   &   &   &  \\
\hline

94 \AA & 0 s      &  11.76$\pm$1.15 s     &  10.9$\pm$1.09 s &  23.15$\pm$1.56 s    & 29.02$\pm$2.38 s & 36.63$\pm$1.97 s & 41.89$\pm$2.84 s\\

%3 (1st peak) & -/-  & -/$\sim$21:45  & $\sim$22:47  & $\sim$21:48  & $\sim$21:48  \\
%3  & $\sim$21:59  & $\sim$22:09  & $\sim$22:19  & $\sim$22:20  & $\sim$22:20  \\
\hline\hline
\end{tabular}}
\label{Table1:time_lag_info}
\end{table}
%%%%%%%%%%%%%%%%%%%%%%%%%%%%%%

%%%%%%%%%%%%%%%%%%%%%%%%%%%%%%%%%%%%
%%% Table 2
%%%%%%%%%%%%%%%%%%%%%%%%%%%%%%%%
\begin{table}[htbp]
\centering
\caption{Time lags between AIA 304~{\AA}, He~\textsc{i}~10830~{\AA} and H$\alpha$~6563~{\AA} passbands }
\setlength{\tabcolsep}{0.9mm}{ 
\begin{tabular}{c | ccc}
\hline \hline
&304 \AA &He~\textsc{i}~10830 \AA &H$\alpha$~6563 \AA   \\
% &   &   &   &   &  \\
\hline

304 \AA & 0 s      &  12.79$\pm$0.37 s     &     35.24$\pm$1.57s   \\

%3 (1st peak) & -/-  & -/$\sim$21:45  & $\sim$22:47  & $\sim$21:48  & $\sim$21:48  \\
%3  & $\sim$21:59  & $\sim$22:09  & $\sim$22:19  & $\sim$22:20  & $\sim$22:20  \\
\hline\hline
\end{tabular}}
\label{Table2:time_lag_info}
\end{table}
%%%%%%%%%%%%%%%%%%%%%%%%%%%%%%%%%%%%

%%%%%%%%%%%%%%%%%%%%%%%%%%%%%%%%%%%%
\subsection{\ya{Variations of the period and magnetic inclination}} 
\label{subsec:frequency_drift_analysis}
%%%%%%%%%%%%%%%%%%%%%%%%%%%%%%%%%%%%%%%%%%%%%%%%%%%

\begin{figure*}
\begin{center}
% trim=left bottom right top
\includegraphics[trim=3cm -1cm 1cm 1cm, width=0.95\textwidth]{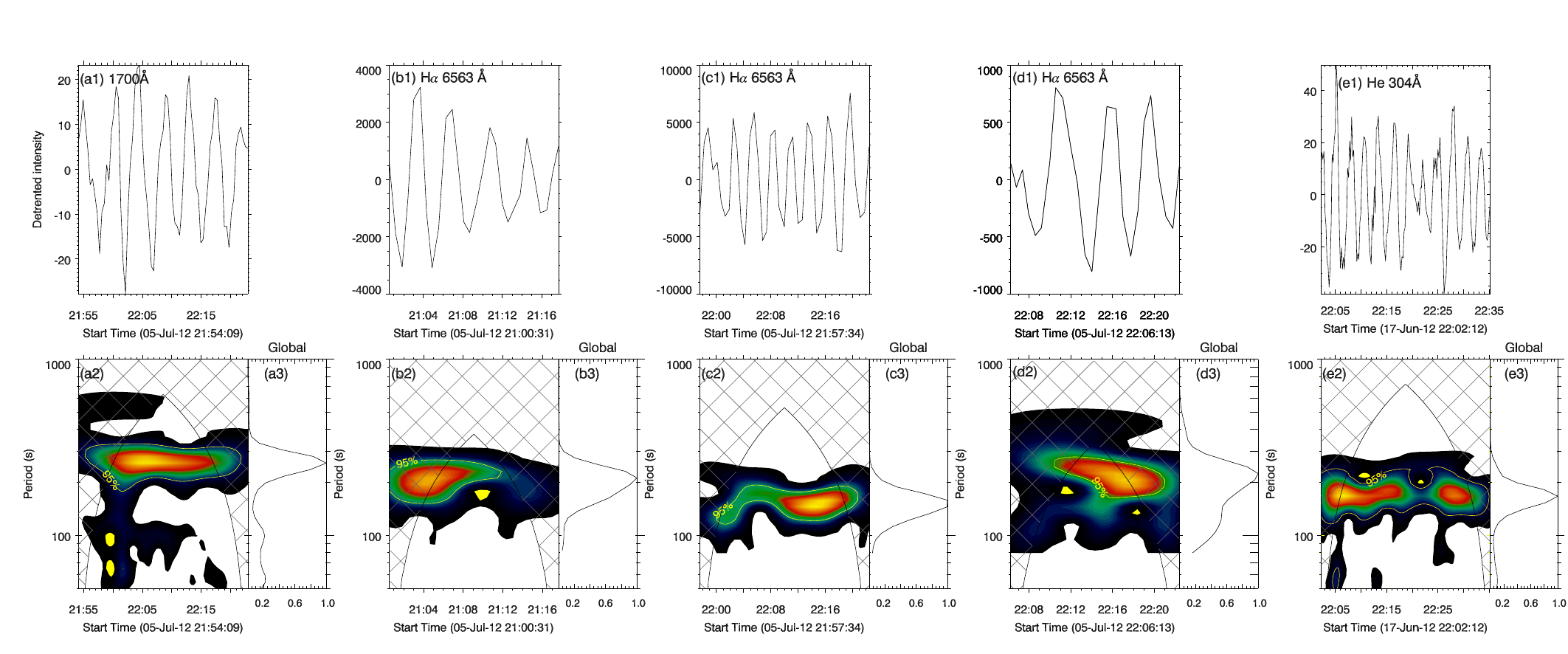}
\caption{\smm{Top panels: the detrended component for the integrated intensity profiles obtained for AIA 1700~{\AA} (panel (a1), during the entire event), H$\alpha$ 6563~{\AA} (panels b1-c1, before and after the flare \ya{at the boundary of umbra and penumbra, and panel d1, after the flare, at the region where the running penumbral wave is still observed),} and AIA 304~{\AA}
(panel e1, after the flare) passbands. These profiles are obtained for the regions shown as white, red and blue solid lines in the time-distance plots shown in Fig. \ref{fig:fig3} i.e over the region between the two red solid lines in AIA 1700~{\AA} (see Fig.~\ref{fig:fig3} panel (c) between 21:51 and 22:20~UT), \ya{red and blue solid lines in }H$\alpha$ 6563~{\AA} \ya{(see Fig.~\ref{fig:fig3} panel (b))} and \ya{white solid lines in} AIA 304~{\AA} \ya{(see Fig.~\ref{fig:fig3} panel (a))}.} Bottom panels: \ya{(a2)-(e2) Morlet wavelet power spectrum and (a3)-(e3) Global wavelet power spectrum for AIA 304, 1700~{\AA} and H$\alpha$ 6563~{\AA} passbands in different stages during the flare. }
\label{fig:fig9}}
\end{center}
\end{figure*}

%\lf{[Why are we interested in looking at frequency changes? Give a sentence or two, either here or in the introduction.]} 
\ya{To investigate the effect of the flare on three-minute oscillation, we studied the frequency changes, which are determined by the angle of the magnetic field to the local vertical and temperature evolution, before and after the flare. The two factors can be affected by a flare.} \smm{We highlighted the regions where the oscillation occurs by solid red, white, and blue horizontal lines in the time-distance plots in Fig.~\ref{fig:fig3}. We created the integrated intensity profiles (see Fig.~\ref{fig:fig9})} for them i.e. between the two \lf{horizontal} \smm{solid} white lines \smm{(see panel (a), between 22:00 and 22:35~UT)} for AIA 304~{\AA}, two red lines for H$\alpha$ before \smm{(see Fig.~\ref{fig:fig3} panel (b) between 21:00 and 21:18~UT)} and after \smm{(see Fig.~\ref{fig:fig3} panel (b) between 22:00 and 22:21~UT)} the flare, \ya{two blue lines for H$\alpha$ after the flare (see Fig.~\ref{fig:fig3} panel (b) between 22:06 and 22:22 UT)} and red lines \smm{(see Fig.~\ref{fig:fig3} panel (c) between 21:51 and 22:20~UT)} for 1700~{\AA}. \smm{The detrended components for the integrated intensity profiles and their} corresponding wavelet analysis \smm{are shown in Fig.~\ref{fig:fig9} for H$\alpha$ 6563~{\AA}, AIA 304 and 1700~{\AA} passbands. In Fig.~\ref{fig:fig9} panel \smm{(a2)}, we observed a 4-minute oscillation in the sunspot's penumbra observed in the AIA~1700~\AA\ passband.} In Fig.~\ref{fig:fig9}, panels \smm{(b2)} and \smm{(c2)} display the periodic property before and after the flare in H$\alpha$ 6563~\AA. Before the flare, the period is more than 200~s. However, the period of the oscillation after the flare is 180~s, which is consistent with the period obtained in the He~\textsc{i} 10830~\AA\ (See Fig.~\ref{fig:fig2}) and AIA 304~{\AA} \smm{(see Fig.~\ref{fig:fig9} panel e2)} passbands. The period of the oscillation is slightly shorter than the running penumbral wave at the same locations. The frequency shift is observed at He~\textsc{i} 10830~\AA, \ya{as shown in Fig.~\ref{fig:fig2}, and presented on the integrated intensity profiles of H$\alpha$ 6563 \AA, as shown in Fig.~\ref{fig:fig9}. Besides, we notice that the RPW co-exists with the enhanced oscillation after the flare and the period of the RPW after the flare is the same as before the flare, which is about 200~s, as shown in Fig.~\ref{fig:fig9} panels (d1$-$d3).} 
The periods observed at various wavelengths during the pre-flare and post-flare phases are summarized in Table~\ref{Table4:periods}
\\
%\lf{[refer also to Figure 6 here.]}\\
%%%
\begin{table}[htbp]
\centering
\caption{Periods observed \smm{at various} wavelengths \smm{during} the pre-flare and post-flare phases}

\setlength{\tabcolsep}{2.0mm}{ 
\begin{tabular}{l|cc}
\hline \hline
\smm{Passbands}  &Pre-flare (21:00$-$21:36~UT)  &EUV late phase (21:55$-$22:30~UT)  \\
\hline
\smm{AIA} 1700~{\AA}   & No data      & 240 s     \\
H$\alpha$ 6563~{\AA} & $\sim$200 s      & 180 s  \ya{and $\sim$200 s}  \\
%\smm{AIA} 304~{\AA}   & Invisible    & 180 s     \\
\smm{He~\textsc{i}}~10830~{\AA}   & 300 s (early pre-flare)   & 180 s   \\
\smm{He~\textsc{i}}~10830~{\AA}   & $\sim$200 s (late pre-flare) & \\
\smm{All EUV passbands}  & Invisible     & 180 s    \\
\hline\hline
\end{tabular}\label{Table4:periods}}
\end{table}
%%%%%%%%%%%%%%%%%%%%%%%%%%%%%%%%%%%%

%%%%%%%%%%%%%%%%%%%%%%%%%%%%%%%%%%%%%%%%%%%%%%%%%%%%%%% 
% Figure 6
%%%%%%%%%%%%%%%%%%%%%%%%%%%%%%%%%%%%%%%%%%%%%%%%%%%%%%% 
%%%%%%%%%%%%%%%%%%%%%%%%%%%%%%%%%%%%%%%%%%%%%%%%%%%%%%%

The three-minute oscillation is an intrinsic property determined by the acoustic cut-off frequency in the chromospheric resonant cavity. \lf{No signal should propagate below} the cut-off frequency $\omega_c$. In an isothermal atmosphere, this is given by 

\begin{equation}
\centering
\omega_{c} = \frac{\gamma g}{2c_{s}}  =  \sqrt{\frac{\gamma\mu g^{2}}{4RT}},
\end{equation}

%$$
%\omega_{c} = \frac{\gamma g}{2c_{s}}  =  \sqrt{\frac{\gamma\mu g^{2}}{4RT}}
%$$
%%%%%%%%%%%%%%%%%%%%%%%%%%%%%%%%%%%%%%%%%%%%%%%%%%%%%%% 
% Figure 10
%%%%%%%%%%%%%%%%%%%%%%%%%%%%%%%%%%%%%%%%%%%%%%%%%%%%%%% 
\begin{figure*}
\begin{center}
% trim=left bottom right top
\includegraphics[trim=1.3cm 0cm 0.5cm 0.9cm, width=0.6\textwidth]{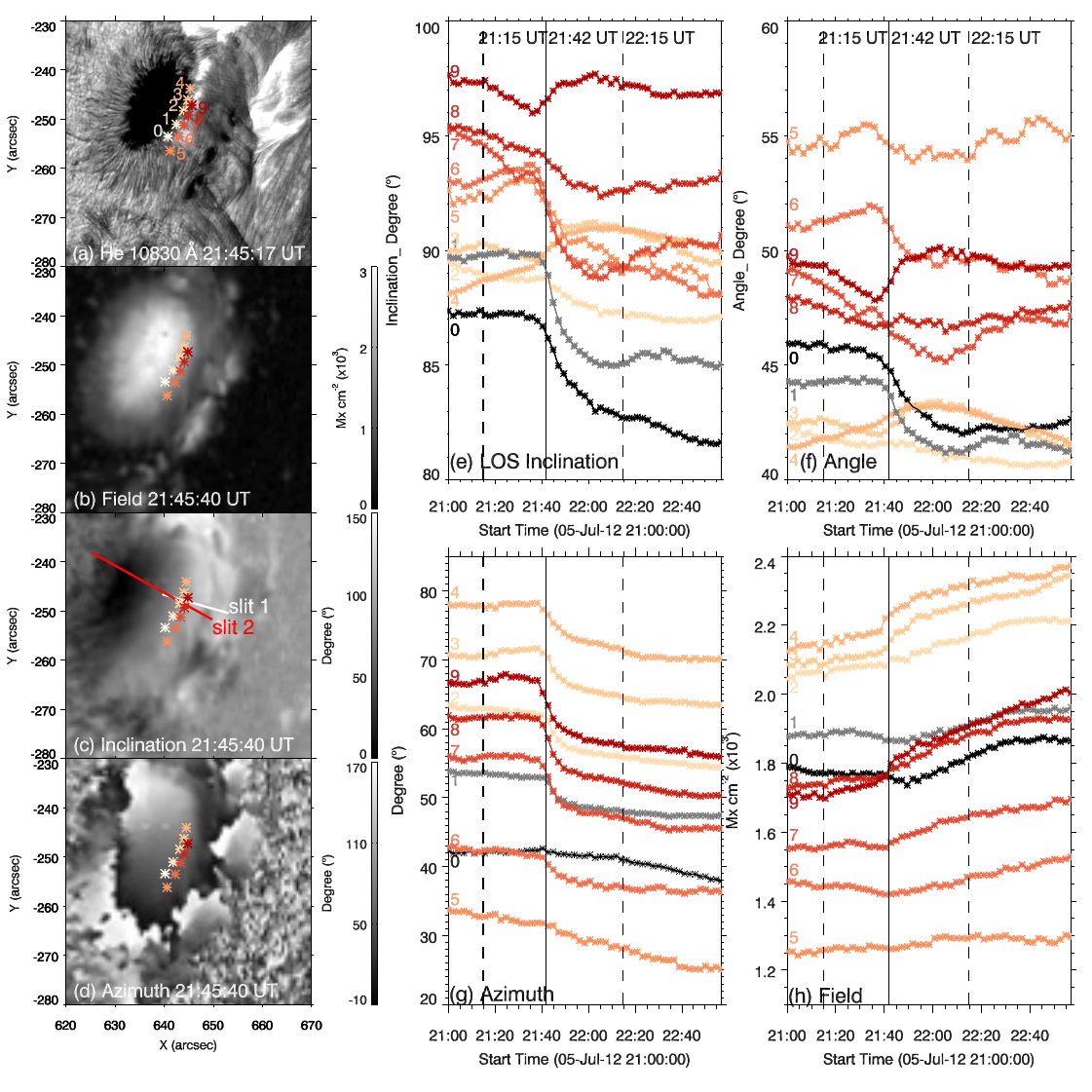}
\caption{\smm{Left panel: Panel (a) shows the He~\textsc{I} 10830~{\AA} image at 21:45~UT. Panel (b$-$d) show the HMI light of sight field strength, inclination and azimuth of the magnetic field respectively. Various points in and around sunspot umbral and penumbral boundaries are shown in coloured asterisks. Right panel: Panel (e) shows the time profile of the inclination of the magnetic field along the line of sight obtained at ten locations that are shown in the left panel. Panel (f) shows the time profile of the angle to the local vertical for a subset of these. Panels (g) and (h) show the time profiles of the azimuth and field strength of the magnetic field along the line of sight at the same ten locations. The two black dashed vertical lines are before the flare at 21:15~UT and after the flare at 22:15~UT where the EM distribution and temperature were measured in Figs.~\ref{fig:fig7} and~\ref{fig:fig8} respectively. The black solid vertical line indicates the flare's peak time, 21:42~UT. The angles to the local vertical for the ten positions at 21:15 and 22:15 are listed in Table~\ref{Table3:parameters}.} \label{fig:fig10}}
\end{center}
\end{figure*}

%In panel (e), to avoid entangled light curves, we only selected 6 points to plot their time profiles.
%%%%%%%%%%%%%%%%%%%%%%%%%%%%%%%%%%%%%%%%%%%%%%%%%%%%%%%

where $\gamma$ is the adiabatic index, $\mu$ is the mean molecular mass, $g$ is the gravitational acceleration, $R$ is the gas constant, and $T$ is the temperature. The chromosphere would filter out frequencies less than the cut-off from the photosphere below \citep{horace1909theory}. However, in magneto-acoustic-gravity (MAG) waves, the acoustic cut-off frequency \lf{that} determines the period of the oscillations \lf{that can propagate is given by}:

\begin{equation}
\centering
\omega_{c} = \frac{\gamma g \cos \theta}{2c_{s}} \varpropto  \frac{g \cos\theta}{\sqrt{T}}
\end{equation}

%$$
%\tau_{cooling} = \tau_{c} [(\frac{\tau_{r}}{\tau_{c}})^{7/12} - 1] + \frac{2}{3} \tau_{r} (\frac{\tau_{c}}{\tau_{r}})^{5/12}  [1 - (\frac{T_L}{T_*})]
%$$

where $\theta$ is the angle to the local vertical, and $T$ represents plasma temperature. The frequency is proportional to the cosine of this angle and inversely proportional to the square root of the temperature. \lf{We assume that $\gamma$ remains constant.} 

The variation of the magnetic structure, therefore, plays an important role in the oscillation. Thus, we studied the variation of magnetic components including magnetic field inclination from the line of sight, azimuth, and \lf{strength} from 21:00 to 23:00~UT. \smm{Photospheric magnetograms are obtained from the Helioseismic and Magnetic Imager \smm{ \cite[HMI; ][]{2012SoPh..275..229S}} on board SDO. The hmi.b$_{-}$135s data series provides full-disk Milne-Eddington inversion\footnote{http://jsoc.stanford.edu/HMI/Vector\_products.html} \citep{2014SoPh..289.3483H} with the magnetic field  strength, inclination, azimuth, disambiguation, etc., information, \lf{at} a cadence of 135~s (2~minutes 15~sec).} 
%The inclination of the magnetic field \lf{is taken} with respect to the line of sight. 
The magnetic field strength has units Mx/cm$^{2}$. 

\smm{In Fig.~\ref{fig:fig10}, panel (a), we showed the He~\textsc{i} image at 21:45~UT along with ten locations (shown as asterisks) on the umbral-penumbral boundary of the sunspot. These locations indicate the regions on the oscillating stripe of the flare ribbon. The maps of the magnetic field strength, inclination and azimuth} are shown in Fig.~\ref{fig:fig10} (panel (b--d)), and we showed the time series of the magnetic field inclination and the angle to the local vertical at these ten locations in panels (e) and (f) respectively. The time series of the magnetic field azimuth and field strength are shown in panels (g) and (h). The vertical solid black line indicates the flare peak time at around 21:42~UT. The two black dashed lines indicate the \lf{times 21:15 and 22:15~UT at which} we studied their emission measure distributions, (see Section~\ref{subsec:temperature_evolution}, Figs.~\ref{fig:fig7} and~\ref{fig:fig8}). %\lf{The interpretation of this is that} the lifting of the EUV late-phase loop led to a decrease in the inclination \lf{[I think that this needs more explanation. We mention the late-phase loops above but we have not yet discussed their behaviour. Perhaps we have to refer to a previous paper? Or explain what we mean by lifting, and why we think it is the actual movement of the loops than just evolution in temperature leading to the appearance of movement.]}. 

 \lf{Over an interval of 10$-$20 minutes,} the inclination from the line-of-sight of the photospheric magnetic field on the oscillating stripe decreases by up to 5$^{\circ}$ at locations 0$-$2 and 5$-$8, indicating that the magnetic field becomes more vertical. \lf{It is reasonable to assume that the chromospheric magnetic field also becomes more vertical.} On the contrary, the inclination at locations 3, 4, and 9 does not decrease. These three locations are possibly at the footpoint of the post-flare loop caused by the magnetic implosion or shrinkage of the post-flare loop. As seen \smm{in panel~(a),} locations 3, 4, and 9 are much closer to the footpoints of the post-flare loop, while the others are at the footpoints of the EUV late-phase loop, \ya{which could become more vertical}.% which gradually rises. 

\ya{In addition, we calculated the angle between the magnetic field and the surface normal and the results are shown in Fig.~\ref{fig:fig10} panel (f). These are further used for the estimation of the frequency ratio and the results are discussed in section~\ref{frequency_ratio_measurement}. The 180$^{\circ}$ ambiguity of azimuth is solved \citep{2014SoPh..289.3483H}. The HMI vector field with components of field strength, inclination and azimuth is converted into spherical coordinate components of Bp, Bt and Br \citep{2013arXiv1309.2392S}. Thus, the angle to the local vertical can be obtained (panel (f)). }In the process of coordinate conversion, the magnetic field components (Bp, Bt, and Br) are dependent on the inclination, azimuth, and field strength. So, their time profiles are shown in Fig.~\ref{fig:fig10} as references.\\

%The reason for the sharp variation of the angle could be caused by the variation of azimuth or the measurement error of the transverse field.
%%%%%%%%%%%%%%%%%%%%%%%%%%%%%%%%%%%%
\subsection{Temperature evolution}
\label{subsec:temperature_evolution}
\begin{figure}
\begin{center}
% trim=left bottom right top
\includegraphics[trim=1.0cm 0cm 0.5cm 0.9cm, width=1.0\textwidth]{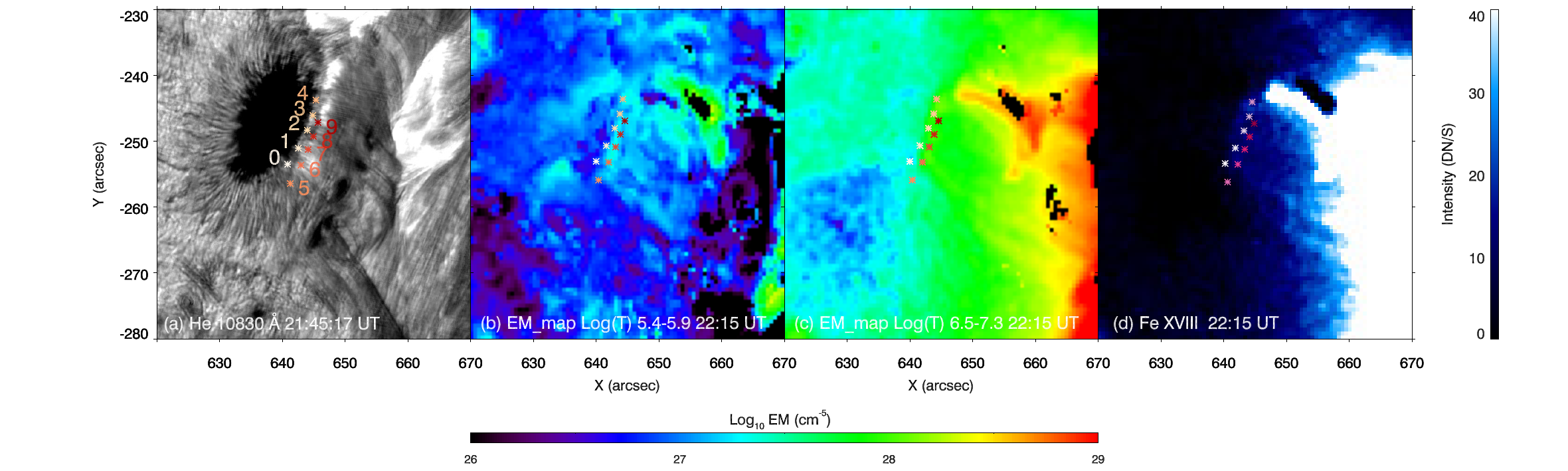}
\caption{\smm{Panel (a): the He~\textsc{i} 10830~\AA\ \smm{image} at 21:45:17~UT; Panels (b$-$c): the emission measure maps created at 22:15~UT in the temperature \smm{range} of log~\smm{\textit{T} [K]} $\sim$ 5.4$-$5.9 and 6.5$-$7.3.} The coloured asterisks in all the panels indicate the ten locations \smm{where we obtained emission measure profiles as a function of temperature shown in Fig.~\ref{fig:fig7} and~\ref{fig:fig8}}. The black patches (at X = 655 to 658\arcsec and Y = -243 to -246\arcsec) in these panels are caused by desaturation. Panel (d): Fe~\textsc{xviii} emission map created at 22:15~UT.} \label{fig:fig6}
\end{center}
\end{figure}

\begin{figure*}
\begin{center}
% trim=left bottom right top
\includegraphics[trim=1.3cm 0cm 0.5cm 0.9cm, width=0.7\textwidth]{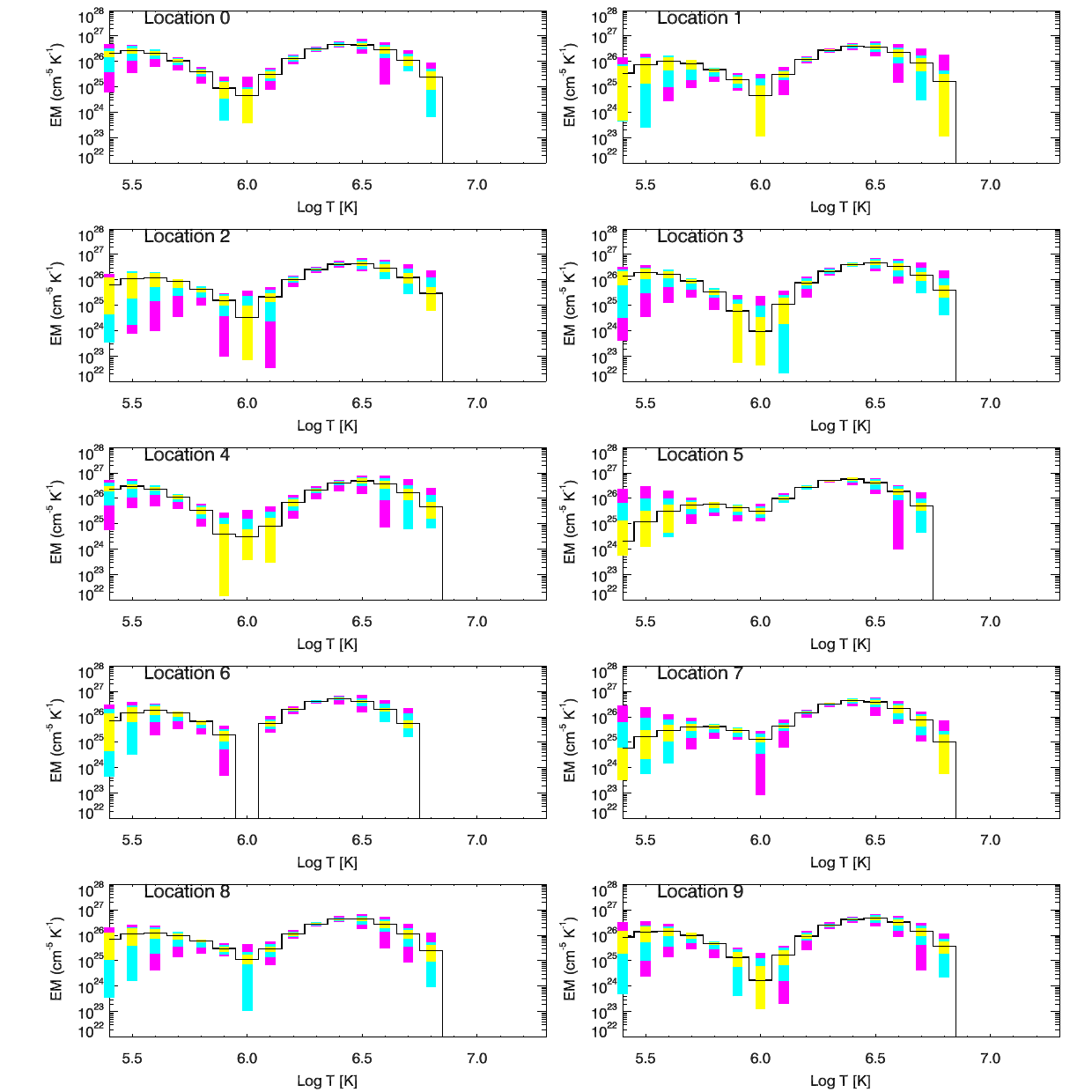}
\caption{EM \smm{profiles at 21:15~UT ($\sim$27 min before the peak of the flare)} obtained at the ten locations (0-9 shown as asterisks in Fig.~\ref{fig:fig6}). The black profiles show the best-fitted EM curves. \smm{The yellow, turquoise and pink coloured bars indicate 50\%, 80\% and 95\% uncertainties associated with EM measurements that are obtained from MC solutions.} \label{fig:fig7}}
\end{center}
\end{figure*}

\begin{figure*}
\begin{center}
% trim=left bottom right top
\includegraphics[trim=1.3cm 0cm 0.5cm 0.9cm, width=0.7\textwidth]{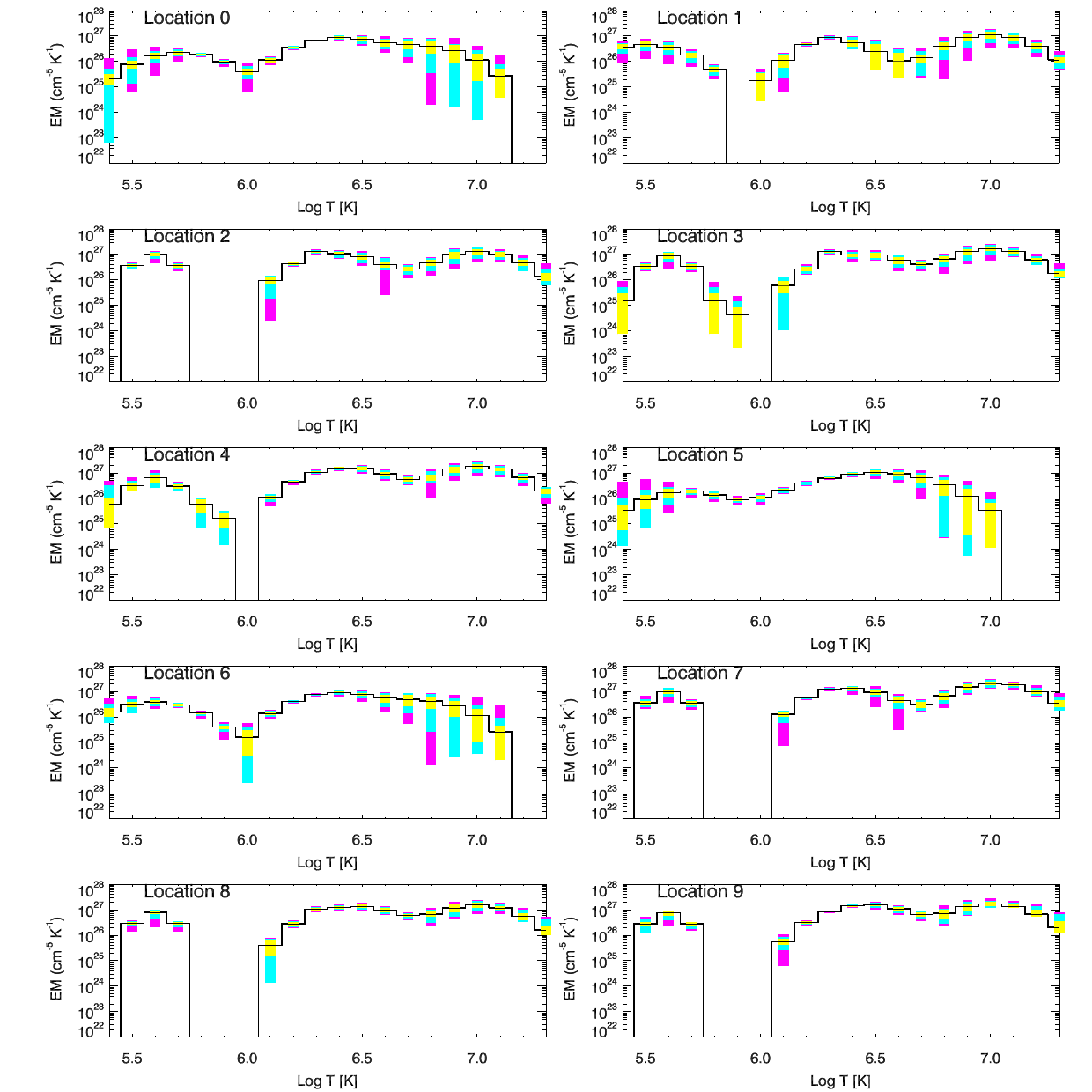}
\caption{EM \smm{profiles at the late phase of the flare at 22:15~UT at the ten locations (0-9 shown as asterisks in Fig.~\ref{fig:fig6})}. The black profiles show the best-fitted EM curves. \smm{The yellow, turquoise and pink coloured bars indicate 50\%, 80\% and 95\% uncertainties associated with EM measurements that are obtained from MC solutions.}  \label{fig:fig8}}
\end{center}
\end{figure*}

%%%%%%%%%%%%%%%%%%%%%%%%%%%%%%%%%%%%
\lf{As mentioned in \smm{section}~\ref{sec1:introduction}, the acoustic cutoff frequency is affected by the chromospheric temperature. We do not have chromospheric temperature measurements, but we can use AIA to look at the temperature evolution at transition region temperatures and above, before and after the flare. This is useful because the chromosphere, transition region, and corona are conductively linked, so this can help us understand whether the temperature structure of the chromosphere might have been changed by the flare.} To obtain the temperature of the plasma at locations where the oscillation occurs, we utilized the emission measure method \citep[EM;][]{2015ApJ...807..143C}. %, Su2018, Li2022
\smm{We used intensities from six AIA coronal passbands (94, 131, 171, 193, 211 and 335~{\AA}) in the EM analysis and obtained the EM distribution over the range of temperatures 5.4$<$log~\textit{T}$<$7.3.} The results are shown in Fig.~\ref{fig:fig6}. In panel (a), \smm{we showed He~\textsc{i} image of the sunspot along with} ten locations (labelled by asterisks) \smm{on the umbral-penumbral boundary of the sunspot}. \smm{We studied} \lf{the} variations of \lf{their} emission measure and temperature \smm{before and after the flare at 21:15 and 22:15~UT respectively (see Fig.~\ref{fig:fig7} and~\ref{fig:fig8})}. To remove the influence of the data saturation \lf{and diffraction fringes} during the impulsive phase and the EUV late phase, we desaturated the AIA data by using the DESAT software package\footnote{https://hesperia.gsfc.nasa.gov/ssw/packages/desat/doc/DESAT\_doc.pdf} available in the SolarSoftware (SSW). \smm{This provides} an automatic de-saturation of AIA images by using a correlation/inversion analysis of the diffraction fringes produced by the telescope. The \lf{resulting} EM maps at \smm{22:15~UT} in the temperature \smm{range, log~\textit{T} [K] $\sim$ 5.4$-$5.9 (0.25$-$0.8~MK) and log~\textit{T} [K] $\sim$ 6.5$-$7.3 (5$-$19~MK) are shown in the panels (b) and (c) of Fig.~\ref{fig:fig6} respectively. The EM values are smaller for points 0$-$4 compared to points 5$-$9 in both temperature ranges. Also, for all points, the EM values are smaller for the EM map in the temperature range of 5.4$-$5.9 compared to the EM map in the temperature range of 6.5$-$7.3.}  
%%%%%%%%%%%%%%%%%%%%%%%%%%%%%%%%%%%%%%%%%%%%%%%%%%%%%%%%%%%%%%%%%
Figs.~\ref{fig:fig7} and~\ref{fig:fig8} \lf{compare} the EM distributions of the ten locations on the oscillating ribbon \smm{27~min} before the flare, at 21:15~UT and \smm{33~min} after the flare, at 22:15~UT \lf{respectively}. The best-fitting EM solution of the observation is indicated by the black curves. \smm{The uncertainties on the EM solutions were measured using 100 Monte Carlo (MC) realizations of the observations. The uncertainty on the AIA intensities was calculated using the function \texttt{aia\_bp\_estimate\_error.pro} in SSW and used in the EM analysis to obtain a 100 EM solution. These solutions are indicated as coloured bars on the best-fit EM solution.}
%, which considers the errors in shot noise, dark subtraction, quantization, read noise, compress noise, photometric calibration, and chianti noise from the CHIANTI model} 

\smm{At 21:15~UT, the EM distributions for all points show two peaks, one at the lower temperature, 5.5$<$log~\textit{T}$<$5.7 (0.3$-$0.5~MK) and the other at high temperature, 6.2$<$log~\textit{T}$<$6.5 (1.6$-$3.2~MK). In the case of the late phase of the flare at 22:15~UT, we observed 3 peaks in the EM distribution. The first two peaks were in the same temperature range as that was observed at 21:15~UT, and the third peak was observed in the very high-temperature range, $\sim$log~\textit{T} [K] = 7. The long error bars in the low, as well as high-temperature part of the EM distribution, are indicative that the DEM at these temperatures is not well constrained. We obtained the EM weighted averaged temperature, \smm{ \textit{\mbox{$\tilde{T}$ = $\int$ EM(T)$\times$T dT / $\int$ EM(T) dT}}} in the entire temperature range 5.4$<$log~\textit{T}$<$7.3 (0.25$-$19~MK). The results are summarized in Table~\ref{Table3:parameters}.} The result shows that the EM in the high-temperature range above \smm{log~\textit{T} [K] = }7 is enhanced at 22:15~UT. As a result, the EM-weighted average temperature also increased. For locations 0, 5, and 6, the emission measures are affected by overlying coronal loops %\lf{[Do we mean `overlying' instead of `nearby'? My recollection of the discussion is that there were overlying loops so it is not possible to isolate the high-temperature emission from the ribbons?]} 
and emission from temperatures above \smm{log~\textit{T} [K] = }7 is \ya{relatively weak}. For the other locations 1$-$4 and 7$-$9, the emission measures in the high-temperature range obviously increase. The footpoints of the late-phase loops at these locations are heated during the EUV late phase.\\

%%%%%%%%%%%%%%%%%%% 
% Figure 7
%%%%%%%%%%%%%%%%%%%%%%%%%%%%%%%%%%%%%%%%%%%%%%%%%%%%%%% 
%%%%%%%%%%%%%%%%%%%%%%%%%%%%%%%%%%%%

%%%%%%%%%%%%%%%%%%%%%%%%%%%%%%%%%%%%%%%%%%%%%%%%%%%%%%% 
% Figure 8
%%%%%%%%%%%%%%%%%%%%%%%%%%%%%%%%%%%%%%%%%%%%%%%%%%%%%%% 
%%%%%%%%%%%%%%%%%%%%%%%%%%%%%%

\smm{In order to find the validity of the high temperature $\sim$7~MK in the EM analysis at 22:15~UT, we carried out further analysis using three AIA channels. As stated earlier in Section~\ref{sec2:overview}, the AIA 94, 171 and 211~{\AA} passbands are dominated by Fe~\textsc{xviii}, Fe~\textsc{x} and Fe~\textsc{xiv} respectively during flares. We measured the contribution of Fe~\textsc{xviii} at 93.93~{\AA} emission in the AIA 94~{\AA} passband using the method by \cite{2013A&A...558A..73D}. This method allows us to subtract the contribution of low temperatures emission in the AIA 94~{\AA} passband by using the low-temperature emission observed in the AIA 211 and 171~{\AA} passbands \( I (Fe~{\textrm{XVIII}}) = I (AIA~94) - I (AIA~211)/120 - I(AIA~171)/450 \). We apply this method to AIA images observed at 22:15~UT and obtained the Fe~\textsc{xviii} emission map which is shown in Fig.~\ref{fig:fig6}, panel (d). This indicates there is a small amount of Fe~\textsc{xviii} emission available at these locations. Kindly note that the contribution function for Fe~\textsc{xviii} line peaks at log~\textit{T} [K] = 6.85 ($\sim$7~MK) (indicating the temperature of formation) but the line could form at a wide range of temperatures, log~\textit{T} [K] = 6.25 to 7.38 (2$-$23~MK). This does not indicate the plasma temperature we observed at 22:15~UT is $\sim$7~MK but it could be between 2 and 7~MK considering that the EM values at high temperatures are not well constrained by the AIA channels. 
}

\subsection{Estimation of the frequency ratio} \label{frequency_ratio_measurement}
%%%%%%%%%%%%%%%%%%%%%%%%%%%%%

\ya{To characterise the frequency change before and after the flare, we estimate the frequency ratios for different EM-weighted averaged temperatures in three different temperature bands. They are 5.4$<$log~\textit{T}$<$7.3 (log~\textit{$\tilde{T}_{total}$}) for the whole temperature range, 5.4$<$log~\textit{T}$<$6.8 (log~\textit{$\tilde{T}_{total2}$}), in which we remove the temperature above log~\textit{T} [K] = 7.0, and lower temperature band, band~I, 5.4$<$log~\textit{T}$<$5.9 (log~\textit{$\tilde{T}_{band~I}$}).} \ya{The frequency ratios before and after the flare are estimated by using equation (2)}. In Table~\ref{Table3:parameters}, the angle to the local vertical of the \smm{photospheric} magnetic field and \smm{EM-weighted averaged temperatures (\textit{$\tilde{T}$)} obtained at the ten locations in the temperature ranges 5.4$<$log~\textit{T}$<$7.3 (log~\textit{$\tilde{T}_{total}$}), 5.4$<$log~\textit{T}$<$6.8 (log~\textit{$\tilde{T}_{total2}$}), band~I, 5.4$<$log~\textit{T}$<$5.9 (log~\textit{$\tilde{T}_{band~I}$}), and  frequency ratios are summarized.}
%at 21:15 and 22:15~UT are listed. Here, we note that there are two temperatures given $-$ the \smm{EM} weighted-average temperature (\ya{log~\textit{$\tilde{T}_{total}$} [K]}) in the whole temperature range and the \lf{EM weighted-average temperature  (\ya{log~\textit{$\tilde{T}_{band I}$} [K]}) , weighted over the} range of \smm{log~\textit{T} [K] = }5.4$-$5.9 %\lf{[minimum? or the weighted average temperature in the 5.4-5.9 range?]}. 
The angle to the local vertical decrease leads to an increase in \lf{the cut-off} frequency, so the period \lf{of waves that can propagate} decreases. Inversely, an increase in temperature leads to a decrease in \lf{cut-off} frequency and an increase in the period. Based on the oscillation observed in the H$\alpha$ 6563~\AA\ line, the period decreases from more than 200~s of the penumbral wave to three minutes of the oscillation. \\

%The inclination of the magnetic field plays a key role in the difference. %The estimated frequency ratios before and after the flare impulsive phase, \lf{calculated using the inclination angles and the EM weighted-average temperature in the low-temperature band}, are also listed in Table~\ref{Table3:parameters}. 
%The estimated frequency ratios by using the lower temperature band (band~I) \ya{and 5.4$<$log~\textit{T}$<$6.8 (log~\textit{$\tilde{T}_{total2}$})} 
The estimated frequency ratios by using the low temperature range are more consistent with the observation compared to the weighted-average temperature in the whole temperature range.  %\lf{[Perhaps we need to explain this a bit more. The temperature that we measure is not the relevant temperature for calculating the acoustic cut-off. That should be the chromospheric temperature. However, we can use the change in transition-region temperature to argue that the change in chromospheric temperature is probably small, so the magnetic effect dominates.]}
\ya{We note that the temperature which determines the acoustic cutoff frequency should be the temperature in the chromosphere or temperature minimum region} \lf{which we cannot measure}. \ya{However, the temperature changes, especially in the low-temperature band or in the transition region are small. Therefore, probably, the chromospheric temperature, which is affected by the temperature change in the transition region} \lf{during the late, conduction-dominated phase of the flare}, \ya{changes very little. }\\\

\begin{table}[!hbtp]
  \centering
 \caption{Angle to the local vertical of the \smm{photospheric} magnetic field, temperature \smm{from the EM distribution} and estimated frequency ratios at ten locations (\smm{that shown in Fig~\ref{fig:fig10}}.)} %\lf{Values are given for properties before (top section) and after (middle section) the flare impulsive phase. The bottom section gives the calculated ratio of the oscillation frequencies before and after the flare, using the temperatures. %[See also my suggestion for how to name the two temperatures]}
 \resizebox{18.8cm}{!} {
 \begin{tabular}{l c c c c c c c c c c c}

\hline
Time (UT) &Location number  &0  &1 &2 & 3 & 4 & 5 & 6 & 7 & 8 & 9 \\
% &   &   &   &   &  \\
\hline

\multirow{3}{*}{21:15} & Angle to the local vertical ($^
{\circ}$)  & 45.74      & 44.26     & 41.72     & 42.57 & 41.79 & 54.54 & 51.27   & 48.35 & 47.40 & 49.19 \\

&\smm{log~\textit{$\tilde{T}_{total}$} (5.4$<$log~\textit{T} $<$7.3)} & 6.18      & 6.30 &     6.33     &  6.27 & 6.16 & 6.25 & 6.20  & 6.29 & 6.28 & 6.26 \\

&\smm{log~\textit{$\tilde{T}_{total2}$} (5.4$<$log~\textit{T} $<$6.8)}
& 6.18       & 6.30    &      6.32   &  6.26 & 6.16 & 6.25 & 6.20  & 6.29 & 6.27 & 6.26\\
 
&\smm{log~\textit{$\tilde{T}_{band~I}$} (5.4$<$log~\textit{T} $<$5.9)} & 5.55     & 5.67    &    5.64   & 5.57 & 5.54& 5.66 & 5.60   & 5.68 & 5.64 &  5.61 \\
%3 (1st peak) & -/-  & -/$\sim$21:45  & $\sim$22:47  & $\sim$21:48  & $\sim$21:48  \\
%3  & $\sim$21:59  & $\sim$22:09  & $\sim$22:19  & $\sim$22:20  & $\sim$22:20  \\
\hline
\multirow{4}{*}{22:15} & Angle to the local vertical ($^
{\circ}$) & 42.18      & 41.47  & 40.86  & 42.98     &42.87  &  54.37  & 49.64  & 45.99   & 46.86 & 49.75 \\

%Weighted-average temperature (log~\textit{T} [K]) 
&\smm{log~\textit{$\tilde{T}_{total}$} (5.4$<$log~\textit{T} $<$7.3)}
& 6.38      & 6.51    &      6.53    &  6.60 & 6.58 & 6.36 & 6.29  & 6.63 & 6.63 & 6.66\\

&\smm{log~\textit{$\tilde{T}_{total2}$} (5.4$<$log~\textit{T} $<$6.8)}
& 6.33       & 6.16    &      6.23    &  6.27 & 6.31 & 6.35 & 6.23  & 6.24 & 6.34 & 6.34\\

&\smm{log~\textit{$\tilde{T}_{band~I}$} (5.4$<$log~\textit{T} $<$5.9)}   & 5.69      & 5.54    &    5.60   & 5.60 & 5.59 & 5.67 & 5.59   & 5.59 & 5.60 &  5.57 \\

\hline
&Frequency ratio (using log~\textit{$\tilde{T}_{total}$}) & 1.18 & 1.2 &   1.24       & 1.47  & 1.64  & 1.13  &  1.07   & 1.42  &  1.49 & 1.61 \\

&Frequency ratio (using log~\textit{$\tilde{T}_{total2}$}) & 1.12 & 0.81 &   0.88       & 1.02  & 1.22  & 1.11  &  1.01   & 0.91  &  1.07 &  1.11 \\

&Frequency ratio (using log~\textit{$\tilde{T}_{band~I}$}) & 1.10 &0.82 &  0.94   & 1.04  & 1.08  & 1.01  &  0.96   & 0.86  & 0.94  & 0.97\\
%3 (1st peak) & -/-  & -/$\sim$21:45  & $\sim$22:47  & $\sim$21:48  & $\sim$21:48  \\
%3  & $\sim$21:59  & $\sim$22:09  & $\sim$22:19  & $\sim$22:20  & $\sim$22:20  \\

\hline
  \end{tabular} \label{Table3:parameters}
 }
\end{table}
%%%%%%%%%%%%%%%%%%%%%%%%%%%%%%%%%%%%%%
\section{Discussion and Conclusion}
\label{sec4:conclusion}
%%%%%%%%%%%%%%%%%%%%%%%%%%%%%%%%%%%%%%

\smm{A thorough investigation of the M1.8 class flare observed on July 5, 2012, was carried out by us \citep{2016ApJ...833..250W, 2018ApJ...859..148W, 2020ApJ...905..126W} and} this is the fourth part of our analysis, focusing on the enhanced three-minute oscillation above the sunspot. This significant oscillation primarily caught our attention in He~\textsc{i} 10830~\AA\ \smm{images} and is also unambiguously observed in H$\alpha$ 6563~\AA\ line center, H$\alpha$ blue wing (-0.75~\AA), and at EUV wavelengths in SDO/AIA instrument. Combining BBSO and SDO, EM analysis, \ya{wavelet analysis,} and analysis of the magnetic components based on the SDO/HMI data, we summarize the observational findings \ya{and discuss} as follows:  \\

\ya{During the pre-flare and post-flare phases, there are differences in oscillation periods, observed at various wavelengths, as summarized in Table~\ref{Table4:periods}. The He~\textsc{i} 10830 \AA\ emission shows periods of 300~s, $\sim$200~s, and 180~s for the photospheric oscillation due to it being optically thin in the early pre-flare, RPW in the late pre-flare and the enhanced oscillation in post-flare phases, respectively. The H$\alpha$ displays 200 s for the RPW before the flare and 180 s for the enhanced oscillation after the flare.  Additionally, we can still see the faint RPW with a period of 200~s after the flare, which co-exists with the enhanced oscillation. For all the EUV passbands, we observed the enhanced oscillation with a period of 180~s, which is consistent with the period of the enhanced oscillation observed in H$\alpha$, He~\textsc{i} 10830 \AA. \\

The RPW observed in the AIA 1700 \AA\ channel, which is dominated by a continuum, shows a period of 240~s. Due to a lack of data in the AIA 1700 \AA\ channel in the pre-flare stage, we investigated the RPW on the opposite side of the flare-related oscillation in the sunspot. By comparison, the irregular oscillation on the eastern side, related to the flare, is also visible in AIA 1700 \AA. In the same observation, the RPW detected in He~\textsc{i} 10830 \AA\ and H$\alpha$ displays  a period of above 200~s. There is only a slight difference between the two wavelengths. For the scenario supposed by, e.g.,  \citet[]{2007ApJ...671.1005B}, the observed wave propagates along the magnetic field lines as a slow magnetoacoustic wave in the sunspot. The formation height is different for  He~\textsc{i} 10830 \AA\ and H$\alpha$ and the projection effect caused by the inclination of the magnetic field leads to the visually different wave pattern with different periods at different heights.}\\

\ya{As described above, the period of the RPW changes into the 180~s period  of the enhanced oscillation at the same location on the boundary of the umbra and penumbra, observed in He~\textsc{i} 10830 \AA\ and H$\alpha$}. 
%From the observation in H$\alpha$ 6563~{\AA} and He~\textsc{i} 10830~\AA\ passbands, the path of the running penumbral wave turns into a stripe of enhanced oscillations before \lf{[?]}and after the impulsive phase of the flare. 
The \lf{period} of the RPW in the time range 21:15$-$21:33~UT is more than 200~s, and its travelling speed is about 14.7$-$27.8~{$\rm km\,s^{-1}$}.  Nevertheless, the enhanced three-minute oscillation at the footpoints of the late-phase loop between 21:55$-$22:20~UT has a shorter period, of three minutes. In the chromospheric cavity, the magneto-acoustic cutoff frequency variation is determined by both the angle from the local vertical of the magnetic field and the temperature. \ya{By comparing the angle to the local vertical of the magnetic field at 21:11~UT and 22:15~UT, we find that} the angle to the local vertical of the magnetic field decreases at the footpoint of the late-phase loops, for seven out of ten locations. With EM analysis, we find that the EM-weighted temperature of the transition region in the temperature range 5.4$<$log~\textit{T}$<$5.9 changes a little \ya{at 21:15 UT and 22:15 UT}. \lf{We do not have a direct measurement of the chromospheric temperature profile, but we can assume that the chromospheric lines are always formed at roughly the same temperature, and the weighted-average temperature in the lowest DEM band barely changes.} So it seems likely that the change in the angle of the field to the local vertical is responsible for the frequency change.\\

The enhanced three-minute oscillation is observed from $\sim$22:00~UT to $\sim$22:35~UT, in the EUV late phase. The oscillation mainly occurs at the footpoints of the late-phase loop, which \lf{is located at} the boundary of the sunspot's umbra and penumbra. During this time, the oscillation is observed in chromospheric lines (He~\textsc{i} 10830 and H$\alpha$ 6563~\AA) and coronal passbands (AIA 94, 131, 171, 193, 211, and 335~\AA) with the same period of around three minutes. \lf{There is a phase lag between passbands, with the oscillation peaking} \ya{first at 94 \AA, then at 131 and 171 \AA, then at 193, 211, 335, and 304 \AA, at last at He~\textsc{i} 10830 and H$\alpha$ passbands.}\\

Recently, there have been studies that suggest activities from a higher layer in the atmosphere can affect the three-minute oscillation above the sunspot, for example, plasma downflows \citep[e.g.,][]{2016ApJ...821L..30K, 2021A&A...645L..12F}. 
\citet{2016ApJ...821L..30K} reported that the weak oscillation is strongly enhanced after the downflow event and suggested that the downflow event drove the three-minute oscillation and caused the associated heating in this region. 
Studies found that an intensity increase occurs in the downflowing phase in the umbral oscillations \citep{2015ApJ...802...45C} and umbral flashes \citep{2017ApJ...845..102H, 2019A&A...627A..46B}. Recent numerical simulations of umbral flashes indicated that downflowing umbral flash is the result of the presence of standing oscillations above sunspot umbra \citep{2021A&A...645L..12F}. They reported a scenario in which a resonant cavity produced by the sharp temperature gradient of the transition region leads to chromospheric standing oscillations.\\

For this study, \lf{it is interesting to identify} whether the oscillation at the footpoint is also caused by standing waves, \lf{in the light of} recent studies revealing that umbral chromospheric waves do not propagate but instead form standing waves \citep{2018A&A...614A..73F}. \ya{In addition, the enhanced oscillation is observed during the EUV late phase when there is additional heating, and downflow is also clearly observed. Considering that we can still observe the faint RPW during almost the same time as the enhanced oscillation, and that the three-minute oscillation is the inherent frequency in the chromosphere, we suggest that the enhanced oscillation with a back-and-forth pattern, in which frequency is determined by the acoustic cutoff frequency in the chromospheric cavity, is affected by the additional heating, maybe related to the downflow. To explore more physical explanations of the enhanced three-minute oscillation and the propagating property, a further simulation study is expected.}\\

\begin{acknowledgments}
We thank the anonymous referee for constructive comments and suggestions. This work is supported by the National Key R\&D program of China 2021YFA1600502 (2021YFA1600500), the Strategic Priority Research Program of the Chinese Academy of Sciences, Grant No. XDB 0560000 (XDB0560102) and NSFC grants 12003072, 12173092, 12073081, 12273115, and 12273101. \smm{SMM and LF acknowledge support from UK Research and Innovation's Science and Technology Facilities Council under grant award numbers ST/T000422/1 and ST/X000990/1.} Y.W. is supported by the China scholarship council and Youth Found of JiangSu No. BK20191108. W. Cao acknowledges support from US NSF grants - AGS-2309939, AGS-1821294 and AST-2108235. We thank Dr. Jianping Li for helpful discussion. We thank the team of SDO/AIA and SDO/HMI for providing valuable data. The AIA and HMI data were downloaded via the Joint Science Operations Center.  \smm{ We gratefully acknowledge the use of data from the Goode Solar Telescope (GST) of the Big Bear Solar Observatory (BBSO). BBSO operation is supported by US NSF AGS-2309939 and AGS-1821294 grants and the New Jersey Institute of Technology. GST operation is partly supported by the Korea Astronomy and Space Science Institute and the Seoul National University. }

\end{acknowledgments}
%%%%%%%%%%%%%%%%%%%%%%%%%%%%%%%%%%%%%%

%%%%%%%%%%%%%%%%%%%%%%%%%%%%%%%%%%%%%%
\bibliography{three_minutes}{}
\bibliographystyle{aasjournal}

\setcounter{figure}{0}
\renewcommand{\thefigure}{A\arabic{figure}}
\appendix

%\section{Appendix information} \label{sec5:appendix}
%\textbf{Based on the last discussion}\\
%Note 1: Figure 11 - The hint of RPW observed in H$\alpha$ in the post-flare phase. The period is about 200 s, the same as pre-flare phase. So, the RPW is still there, which is not replaced by enhanced oscillation.

%%%%%%%%%%%%%%%%%%%%%%%%%%%%%%%%%%

\begin{figure}[htbp]
\begin{center}
% trim=left bottom right top
\includegraphics[trim=0.6cm 0.4cm 0.5cm 0.9cm, width=1.0\textwidth]
{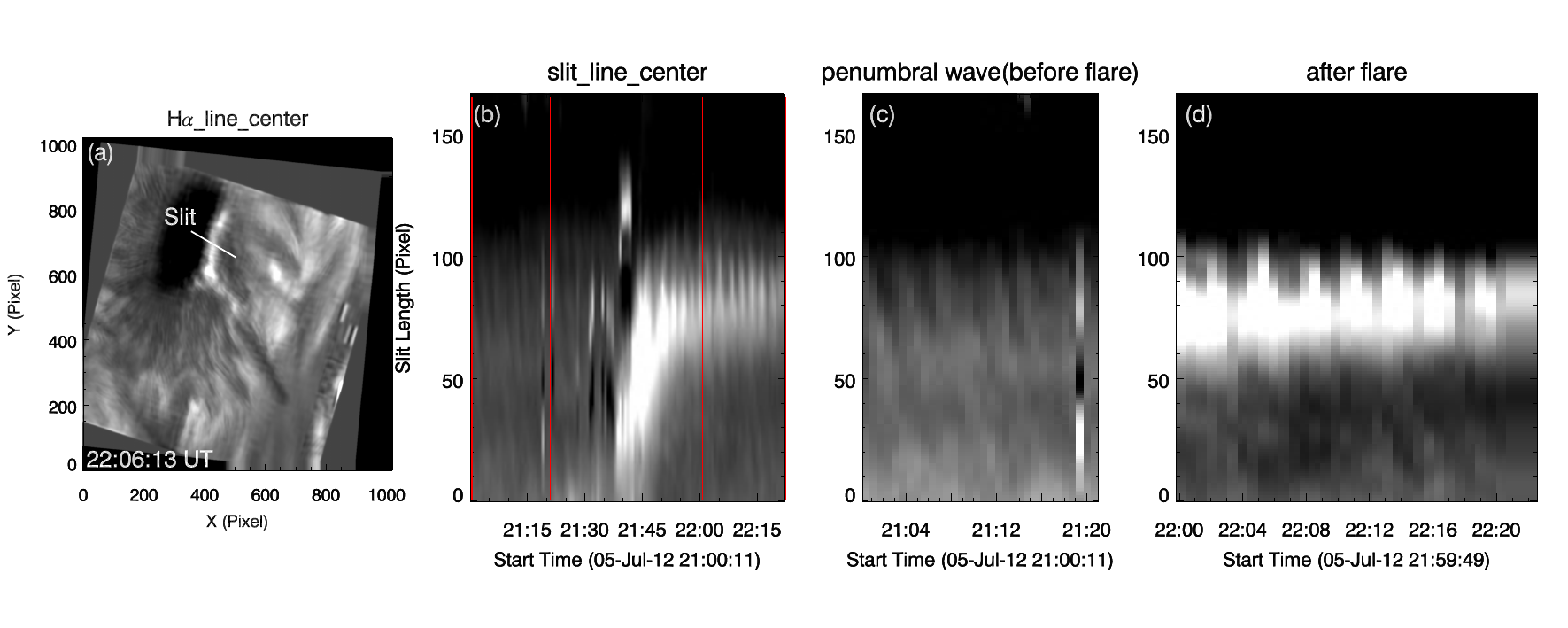}
\caption{The evolution of the sunspot's umbral-penumbral boundary region in H$\alpha$ line center where the enhanced oscillation and penumbral wave coexist during 22:00 and 22:20~UT. Panel (a) shows the H$\alpha$ image at the line center at 22:06:13 UT. The white line indicates the slit position along which we created time-distance plots that are shown in other panels. Panel (b): the oscillation at the boundary of umbra and penumbra over the whole event from 21:00 to 22:20~UT. The regions highlighted with two red boxes indicate the regions shown in panels (c) and (d). Panel (c): The grey and black stripes indicate the penumbral wave before the flare. Panel (d): The coexistence of the enhanced oscillation and penumbral wave after flare from 22:00 to 22:20~UT.  }  \label{fig12}
\end{center}
\end{figure}
%%%%%%%%%%%%%%%%%%%%%%%%%%%%%%

%\begin{figure*}
%\begin{center}
% trim=left bottom right top
%\includegraphics[trim=1.3cm 0cm 0.5cm 0.9cm, width=0.7\textwidth]{img/period_h_alpha_plot_rpw}
%\caption{The RPW observed in the H$\alpha$ labelled by the two red lines (upper panel); bottom left: The detrended lightcurve between the two red lines; bottom right: the wavelet of the detrended lightcurve.}  \label{fig:fig11}}
%\end{center}
%\end{figure*}



\end{document}